\documentclass{egpubl}
\usepackage{pg2020}

\SpecialIssuePaper         

\usepackage[T1]{fontenc}
\usepackage{dfadobe}

\usepackage{cite}  
\BibtexOrBiblatex
\electronicVersion
\PrintedOrElectronic
\ifpdf \usepackage[pdftex]{graphicx} \pdfcompresslevel=9
\else \usepackage[dvips]{graphicx} \fi

\usepackage{egweblnk}

\title[A Bayesian Inference Framework for Procedural Material Parameter Estimation]%
{A Bayesian Inference Framework for \\ Procedural Material Parameter Estimation\\[-0.5cm]}

\author[Y. Guo, M. Ha\v{s}an, L. Yan \& S. Zhao]
{\parbox{\textwidth}{\centering 
		Y. Guo$^1$\orcid{0000-0002-3420-6619}, 
		M. Ha\v{s}an$^2$\orcid{0000-0003-3808-6092},
		L. Yan$^3$ and 
		S. Zhao$^1$\orcid{0000-0003-4759-0514}
		}
	\\
	{\parbox{\textwidth}{\centering
		$^1$University of California, Irvine, USA \hspace{0.2in}
		$^2$Adobe Research, USA \hspace{0.2in}
		$^3$University of California, Santa Barbara, USA	
		}
	}
}

\usepackage{xcolor}

\usepackage{contour}
\contourlength{0.1em}
\usepackage[export]{adjustbox}

\usepackage[linesnumbered,ruled,vlined]{algorithm2e} 

\SetAlFnt{\small}
\SetAlCapFnt{\small}
\SetAlCapNameFnt{\small}
\SetAlCapHSkip{0pt}
\let\oldnl\nl
\newcommand{\nonl}{\renewcommand{\nl}{\let\nl\oldnl}}

\usepackage{amsmath}

\usepackage{amssymb}
\usepackage{booktabs} 
\usepackage{xspace}
\usepackage{paralist}
\usepackage{bm}       
\usepackage{relsize}      
\usepackage{subcaption}
\usepackage{caption}
\usepackage{overpic}
\usepackage{makecell}
\usepackage{multirow}
\usepackage{float}

\newcommand{\Reals}{\mathbb{R}}
\newcommand{\summ}{\mathbb{S}}

\newcommand{\bz}{\bm{z}}
\newcommand{\btheta}{\bm{\theta}}
\newcommand{\target}{\bm{I}_t}
\newcommand{\synth}{\bm{I}_s}
\newcommand{\bsigma}{\bm{\Sigma}}

\newcommand{\pc}{\alpha}

\newcommand{\revision}[1]{\textcolor{black}{#1}}

\newlength{\resultwidth}
\newlength{\resultwidthpdf}
\newcommand{\imglabel}[1]{\put(2,5){\small\bfseries\contour{black}{\textcolor{white}{#1}}}}
\newcommand{\imglabeltop}[1]{\put(2,85){\small\bfseries\contour{black}{\textcolor{white}{#1}}}}

\graphicspath{{images/}}
\begin{document}
 	\teaser{
	\centering
	\vspace{-1cm}
	\addtolength{\tabcolsep}{-4.5pt}
	\newlength{\insetLen}
	\includegraphics[width=0.98\textwidth]{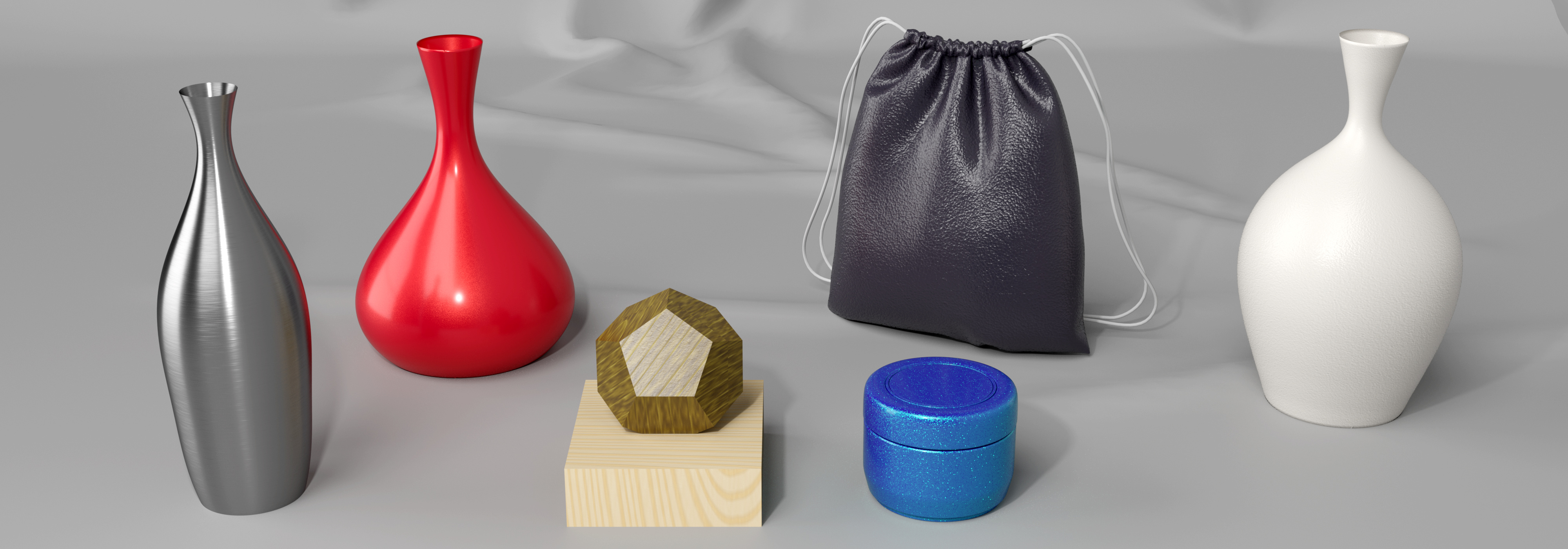}\\[2pt]
	\setlength{\insetLen}{0.132\textwidth}
	\begin{tabular}{cccccccc}
		\raisebox{7pt}{\rotatebox{90}{\small \bfseries Input}} &
		\adjincludegraphics[width=\insetLen,trim={0 {.2\height} 0 {.2\height}},clip]{fig7/5_metal_3/target.jpg} &
		\adjincludegraphics[width=\insetLen,trim={0 {.2\height} 0 {.2\height}},clip]{fig7/4_flake_4/target.jpg} &
		\adjincludegraphics[width=\insetLen,trim={0 {.2\height} 0 {.2\height}},clip]{fig7/6_wood_4/target.jpg} &
		\adjincludegraphics[width=\insetLen,trim={0 {.2\height} 0 {.2\height}},clip]{fig7/6_wood_5/target.jpg} &
		\adjincludegraphics[width=\insetLen,trim={0 {.2\height} 0 {.2\height}},clip]{fig5/4_flake_1/target.jpg} &
		\adjincludegraphics[width=\insetLen,trim={0 {.2\height} 0 {.2\height}},clip]{fig7/2_leather_3/target.jpg} &
		\adjincludegraphics[width=\insetLen,trim={0 {.2\height} 0 {.2\height}},clip]{fig7/1_bump_4/target.jpg}
		\\
		\raisebox{3pt}{\rotatebox{90}{\small \bfseries Rendered}} &
		\adjincludegraphics[width=\insetLen,trim={0 {.2\height} 0 {.2\height}},clip]{fig7/5_metal_3/good1.jpg} &
		\adjincludegraphics[width=\insetLen,trim={0 {.2\height} 0 {.2\height}},clip]{fig7/4_flake_4/good1.jpg} &
		\adjincludegraphics[width=\insetLen,trim={0 {.2\height} 0 {.2\height}},clip]{fig7/6_wood_4/good1.jpg} &
		\adjincludegraphics[width=\insetLen,trim={0 {.2\height} 0 {.2\height}},clip]{fig7/6_wood_5/good1.jpg} &
		\adjincludegraphics[width=\insetLen,trim={0 {.2\height} 0 {.2\height}},clip]{fig5/4_flake_1/good1.jpg} &
		\adjincludegraphics[width=\insetLen,trim={0 {.2\height} 0 {.2\height}},clip]{fig7/2_leather_3/good1.jpg} &
		\adjincludegraphics[width=\insetLen,trim={0 {.2\height} 0 {.2\height}},clip]{fig7/1_bump_4/good1.jpg}
	\end{tabular}
	\captionsetup{labelfont=bf,textfont=it}
	\caption{
			A scene rendered with material parameters estimated using our method: bumpy dielectrics, leather, plaster, wood, brushed metal, and metallic paint. The insets show a few examples of the input (target) images, and renderings produced using our procedural models with parameters found by Bayesian posterior sampling.
		\vspace{3mm}
 	}
	\label{fig:teaser}
}

	\maketitle
	\begin{abstract}
		Procedural material models have been gaining traction in many applications thanks to their flexibility, compactness, and easy editability.
We explore the inverse rendering problem of procedural material parameter estimation from photographs, presenting a unified view of the problem in a Bayesian framework.
In addition to computing point estimates of the parameters by optimization, our framework uses a Markov Chain Monte Carlo approach to sample the space of plausible material parameters, providing a collection of plausible matches that a user can choose from, and efficiently handling both discrete and continuous model parameters.
To demonstrate the effectiveness of our framework, we fit procedural models of a range of materials---wall plaster, leather, wood, anisotropic brushed metals and layered metallic paints---to both synthetic and real target images.

	
	\end{abstract}
	\section{Introduction}
\label{sec:intro}
Physically accurate simulation of material appearance is an important yet challenging problem, with applications in areas from entertainment to product design and architecture visualization.
A key ingredient to photorealistic rendering is high-quality material appearance data.
Acquiring such data from physical measurements such as photographs has been an active research topic in computer vision and graphics. Recently, \emph{procedural} material models have been gaining significant traction in the industry (e.g., Substance \cite{Substance}).
In contrast to traditional texture-based spatially varying BRDFs that represent the variation of surface albedo, roughness, and normal vectors as 2D images, procedural models generate such information using a smaller number of user-facing parameters, providing high compactness, easy editability, and automatic seamless tiling.

The estimation of procedural model parameters faces several challenges. First, the procedural generation and physics-based rendering of materials is a complex process with a diverse set of operations, making the relationship between procedural model parameters and properties of the final renderings non-linear and non-trivial.
Additionally, designing a suitable loss function (metric) to compare a synthesized image to a target image is not obvious. Finally, given the soft nature of the image matching problem, a single point estimate of the ``best'' match may be less informative than a collection of plausible matches that a user can choose from.

In this paper, we introduce a new computational framework to estimate the parameters of procedural material models that focuses on these issues.
Our framework enjoys high generality by not requiring the procedural model to take any specific form, and supporting any differentiable BRDF models, including anisotropy and layering.

To design the loss function, we consider neural summary functions (embeddings) based on Gram matrices of VGG feature maps \cite{Gatys2015,Gatys2016}, as well as hand-crafted summary functions~(\S\ref{sec:summary_func}). The VGG feature map approach is becoming standard practice in computer vision, and was first introduced to material capture by Aittala et al. \cite{Aittala2016}; we extend this approach to procedural material estimation.

\begin{figure*}[t]
	\includegraphics[width=\textwidth]{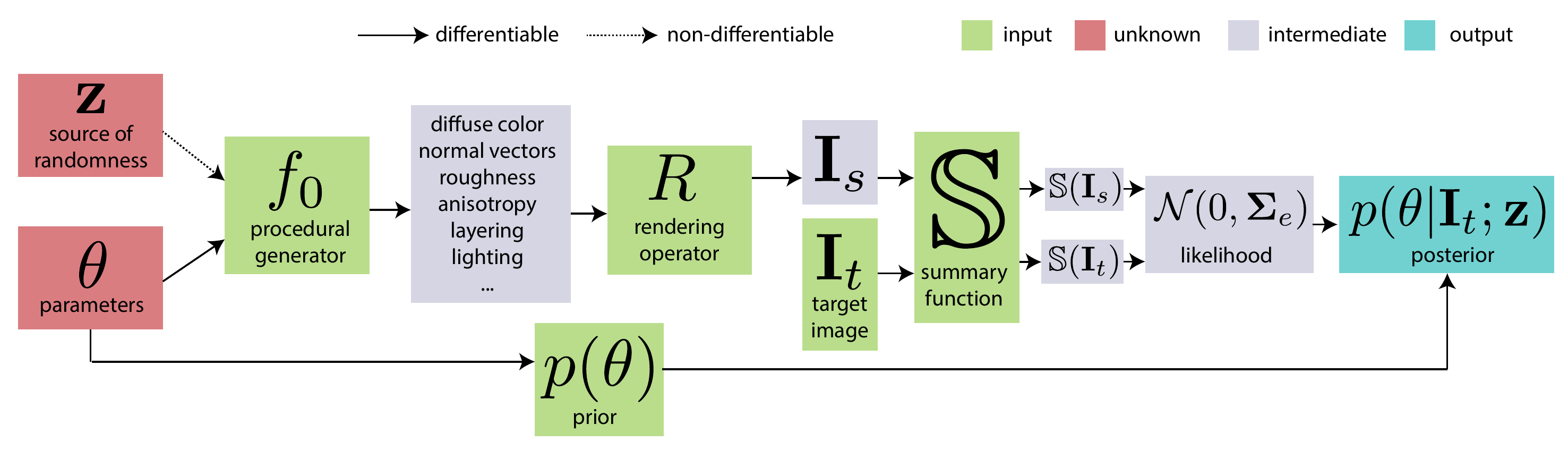}
	\captionsetup{labelfont=bf,textfont=it}
	\caption{Our posterior computation combines priors, a procedural material model, a rendering operator, a summary function, and a target image. This posterior distribution can then be sampled to provide plausible values of the parameter vector. The value of the posterior is computed up to a normalization term, which does not effect MCMC sampling. The entire posterior definition is differentiable in the material parameters (excluding optional discrete model parameters).}
	\label{fig:posterior}
\end{figure*}

We make two main contributions. The first contribution is a unified view of the procedural parameter estimation problem in a \emph{Bayesian framework}~(\S\ref{sec:bayesian}), precisely defining the posterior distribution of the parameters given the captured data and priors, allowing for both maximization and sampling of the posterior. Four components (priors, procedural material model, rendering operator, summary function) together define our posterior distribution (outlined in Figure~\ref{fig:posterior}). 

Our second contribution is to introduce a \emph{Bayesian inference} approach capable of drawing samples from the space of plausible material parameters.
This provides additional information beyond single point estimates of material parameters (for example, though not limited to, discovering similarity structures in the parameter space).
Further, due to an ability to combine multiple Markov-Chain Monte Carlo (MCMC) sampling techniques such as Metropolis-Hasting (MH), Hamiltonian Monte Carlo (HMC), and Metropolis-adjusted Langevin algorithm (MALA), our technique is capable of efficiently handling both discrete and continuous model parameters.
Posterior sampling is a well-studied area within statistics and has been used in computer vision and inverse rendering \cite{Picture}, but to our knowledge, it has not yet been applied to material appearance acquisition.

To demonstrate the effectiveness of our framework, we fit procedural models for a diverse set of materials from standard opaque dielectrics (e.g. plastics, leather, wall paint) to dielectrics with 3D structure (wood) to anisotropic brushed metals and layered metallic paints (see Figure~\ref{fig:teaser}, \S\ref{sec:results}, and the supplemental materials).


	\section{Related Work}
\label{sec:prior_work}
We review previous work on material parameter estimation in computer graphics and vision, as well as on Markov-Chain Monte Carlo (MCMC) methods in Bayesian inference.

\paragraph*{SVBRDF capture.} A large amount of previous work focuses on acquisition of material data from physical measurements. The methods generally observe the material sample with a fixed camera position, and solve for the parameters of a spatially-varying BRDF model such as diffuse albedo, roughness (glossiness) and surface normal. They differ in the number of light patterns required and their type; the patterns used include moving linear light \cite{Gardner2003}, Gray code patterns \cite{Francken2009} and spherical harmonic illumination \cite{Ghosh2009}. In these approaches, the model and its optimization are specific to the light patterns and the optical setup of the method, as general non-linear optimization was historically deemed  inefficient and not robust enough.

More recently, Aittala et al. \cite{Aittala2013} captured per-pixel SVBRDF data using Fourier patterns projected using an LCD screen; their algorithm used a fairly general, differentiable forward evaluation model, which was inverted in a maximum a-posteriori (MAP) framework. Later work by Aittala et al. \cite{Aittala2015,Aittala2016} found per-pixel parameters of stationary spatially-varying SVBRDFs from two-shot and one-shot flash-lit photographs, respectively. In the latter case, the approach used a neural Gram-matrix texture descriptor based on the texture synthesis and feature transfer work of Gatys \cite{Gatys2015,Gatys2016} to compare renderings with similar texture patterns but without pixel alignment. We demonstrate that this descriptor makes an excellent summary function within our framework; in fact, the approach works well in our case, as the procedural nature of the model serves as an additional implicit prior, compared to per-pixel approaches. On the other hand, our forward evaluation process is more complex than Aittala et al., since it also includes the procedural material generation itself.

Recent methods by Deschaintre et al. \cite{Deschaintre2018}, Li et al. \cite{Li2018} have been able to capture non-stationary SVBRDFs from a single flash photograph by training an end-to-end deep convolutional network. Gao et al. \cite{Gao2019} introduced an auto-encoder approach, optimizing the appearance match in the latent space. All of these approaches estimate per-pixel parameters of the microfacet model (diffuse albedo, roughness, normal), and are not obviously applicable to estimation of procedural model parameters, nor to more advanced optical models (significant anisotropy, layering or scattering).

\paragraph*{Procedural material parameter estimation.} Focus on estimating the parameters of procedural models has been relatively rare. The dual-scale glossy parameter estimation work of Wang et al. \cite{Wang2011} finds, under step-edge lighting, the parameters of a bumpy surface model consisting of a heightfield constructed from a Gaussian noise power spectrum and global microfacet material parameters. Their results provide impressive accuracy, but the solution is highly specialized for this material model and illumination.

Recently, Hu et al. \cite{Hu2019} introduced a method for inverse procedural material modeling that treats the material as a black box, and trains a neural network mapping images to parameter vector predictions. The training data comes from evaluating the black box model for random parameters. In our experiments, this approach was less accurate; our fully differentiable models can achieve higher accuracy fits and can be used to explore posterior distributions through sampling. In a sense, this neural prediction method could be seen as orthogonal to ours, as we could use it for initialization of our parameter vector, continuing with our MCMC sampling.

\paragraph*{Optical parameters of fiber-based models.} Several approaches for rendering of fabrics model the material at the microscopic fiber level \cite{Zhao2011,Zhao2016,Leaf2018}. However, the optical properties of the fibers (e.g. roughness, scattering albedo) have to be chosen separately to match real examples. Zhao et al. \cite{Zhao2011} use a simple but effective trick of matching the mean and standard deviation (in RGB) of the pixels in a well-chosen area of the target and simulated image. Khungurn et al. \cite{Khungurn2015} have extended this approach with a differentiable volumetric renderer, combined with a stochastic gradient descent; however, their method is still specific to fiber-level modeling of cloth.

\paragraph*{Bayesian inference and MCMC.} A variety of methods used across the sciences are Bayesian in nature; in this paper, we specifically explore Bayesian inference for parameter estimation through Markov-Chain Monte Carlo (MCMC) sampling of the posterior distribution. Provided a nonnegative function~$f$, MCMC techniques can draw samples from the probability density proportional to the given function~$f$ without knowing the normalization factor. Metropolis-Hastings \cite{Hastings} is one of the most widely used MCMC sampling methods. If $f$ is differentiable, the presence of gradient information leads to more efficient sampling methods such as Hamiltonian Monte Carlo (HMC)~\cite{Neal2012,Betancourt2017} and Metropolis-adjusted Langevin algorithm (MALA)~\cite{MALA}.
Our inference framework is not limited to any specific MCMC sampling technique.
In practice, our implementation handles discrete model parameters using MH and continuous ones using MALA (with preconditioning~\cite{Santa}). We opt MALA for its simpler hyper-parameter tweaking (compared to HMC).


\paragraph*{MCMC applications in graphics and vision.} Markov chain Monte Carlo techniques have been heavily studied in rendering, though not for Bayesian inference, but rather for sampling light transport paths with probability proportional to their importance; notably Metropolis light transport \cite{MLT} and its primary sample space variant \cite{Kelemen}. Much further work has built on these techniques, including more recent work that uses a variant of Hamiltonian Monte Carlo \cite{H2MC}. However, all of these approaches focus on better sampling for traditional rendering, rather than parameter estimation in inverse rendering tasks.

In computer vision, Bayesian inference with MCMC has been used for the inverse problems of scene understanding. A notable previous work is Picture \cite{Picture}, a probabilistic system and programming language for scene understanding tasks, for example (though not limited to) human face and body pose estimation. The programming language is essentially used to specify a forward model (e.g., render a face in a given pose), and the system then handles the MCMC sampling of the posterior distribution through a combination of sampling (proposal) techniques. This is closely related to the overall design of our system. However, the Picture system does not appear to be publicly available, and our application is fairly distant from its original goals.

	\section{Preliminaries}
\label{sec:prelim}

\paragraph*{Procedural model generation.}
We focus on \emph{procedural material models} which utilize specialized 
procedures (pieces of code) to generate spatially varying surface reflectance information.
Specifically, let $\btheta$ be the parameters taken by some procedural material generation process $f_0$.
Then, $f_0(\btheta)$ generates the material properties (e.g., albedo, roughness, surface normals, anisotropy, scattering, etc.), in addition to any other parameters required by rendering (e.g. light parameters), which can in turn be converted into a synthetic image $\synth$ via a rendering operator $R$.
This \emph{forward evaluation} process can be summarized as
\begin{equation}
	\label{eq:forward}
	\synth = R(f_0(\btheta)) = f(\btheta),
\end{equation}
where $f$ is the composition of $R$ and $f_0$.

When modeling real-world materials, it is desirable to capture naturally arising irregularities.
In procedural modeling, this is usually achieved by making the model generation process $f_0$ to take extra random input $\bz$ (e.g., random seeds, pre-generated noise textures, etc.) that is then used to create the irregularities.
This also causes the full forward evaluation to become $f(\btheta; \bz) := R(f_0(\btheta; \bz))$.

\paragraph*{Continuous and discrete parameters.}
While most procedural material parameters tend to be continuous, discrete parameters can be useful for switching certain components on and off, or for choosing between several discrete noise types. We model this by splitting the parameter vector into continuous and discrete components, $\btheta = (\btheta_c, \btheta_d)$.
We assume the forward evaluator $f$ to be differentiable with respect to $\btheta_c$ (but not $\btheta_d$ or the random input $\bz$).

\paragraph*{Inverse problem specification.}
We consider the problem of inferring procedural model parameters $\btheta$ given a target image $\target$  (which is typically a photograph of a material sample under known illumination).
This, essentially, requires inverting $f$ in Eq.~\eqref{eq:forward}: $\btheta = f^{-1}(\target)$. Direct inversion of $f = R \circ f_0$ is intractable for any but the simplest material and rendering models.
Instead, we aim to identify a collection of plausible values $\btheta$ such that $\synth$ has similar appearance to $\target$:
\begin{equation}
	\label{eq:approx}
	\mbox{find examples of } \ \btheta \ \mbox{s.t.} \ \target \approx f(\btheta; \bz),
\end{equation}
for some (any) $\bz$, where $\approx$ is an \emph{appearance-match} relation that will be discussed in the next section.

	\section{Summary Functions}
\label{sec:summary_func}

\begin{figure}[t]
	\addtolength{\tabcolsep}{-5pt}
	\begin{tabular}{cccc}
		\includegraphics[width=0.24\columnwidth]{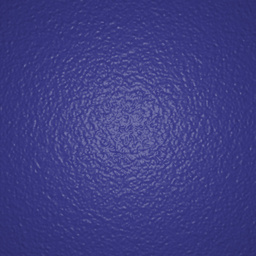} &
		\includegraphics[width=0.24\columnwidth]{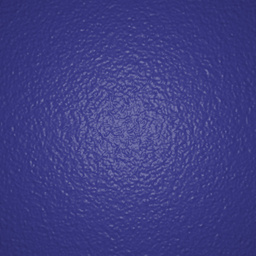} &
		\includegraphics[width=0.24\columnwidth]{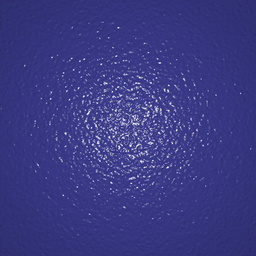} &
		\includegraphics[width=0.24\columnwidth]{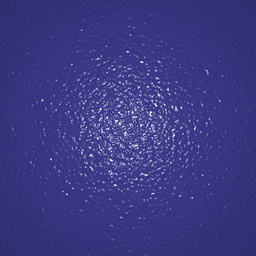} \\
		(a1) & (a2) & (b1) & (b2)
	\end{tabular}
	\captionsetup{labelfont=bf,textfont=it}
	\caption{\label{fig:syn1}
		Each pair of images among (a, b) are generated using identical model parameters $\btheta$ but different irregularities $\bz$. The pixel-wise L2 norm of the difference between these image pairs is large and not useful for estimating model parameters.
	}
\end{figure}

To solve the parameter estimation problem using Eq.~\eqref{eq:approx}, a key ingredient is the appearance-match relation.
Unfortunately, we cannot use simplistic image difference metrics such as the L2 or L1 norms.
This is because the features (bumps, scratches, flakes, yarns, etc.) in the images of real-world materials are generally misaligned, even when the two images represent the same material.
In procedural modeling, as shown in Figure \ref{fig:syn1}, with irregularities created differently using $\bz_1$ and $\bz_2$, the same procedural model parameters $\btheta$ can yield slightly different results $f(\btheta; \bz_1)$ and $f(\btheta, \bz_2)$.

To overcome this challenge, we use the concept of a \emph{summary function}, which abstracts away the unimportant differences in the placement of the features, and summarizes the predicted and target images into smaller vectors whose similarity can be judged with simple metrics like L2 distance.

We define an image summary function $\summ$ to be a continuous function that maps an image of a material ($\target$ or $\synth$) into a vector in $\Reals^k$. An idealized summary function would have the property that
\begin{equation}
	\summ(f(\btheta_1, \bz_1)) = \summ(f(\btheta_2, \bz_2)) \ \Leftrightarrow \ \btheta_1 = \btheta_2.
\end{equation}
That is, applying the summary function would 
abstract away from the randomness $\bz$ and the difference between the two summary vectors would be entirely due to different material properties $\btheta$. Practical summary functions will satisfy the above only approximately. However, a good practical summary function will embed images of the same material close to each other, and images of different materials further away from each other. Below we discuss several techniques for constructing summary functions.

\paragraph*{Neural summary function.}
Gatys et al. \cite{Gatys2015,Gatys2016} introduced the idea of using the features of an image classification neural network (usually VGG \cite{VGG}) as a descriptor $T_G$ of image texture (or style). Optimizing images to minimize the difference in $T_G$ (combined with other constraints) allowed Gatys et al. to produce impressive, state-of-the art results for parametric texture synthesis and style transfer between images. While further work  has introduced improvements \cite{Risser2017}, we find that the original version from Gatys et al. works already well in our case.

Aittala et al. \cite{Aittala2016} introduced this technique to capturing material parameter textures (albedo, roughness, normal and specular maps) of stationary materials. They optimized for a $256 \times 256$ stationary patch that matches the target image in various crops, using a combination of $T_G$ and a number of special Fourier-domain priors. In our case (for procedural materials), we find that the neural summary function $T_G$ works even more effectively; we can simply apply it to the entire target or simulated images (not requiring crops nor Fourier-domain priors).
Specifically, define the Gram matrix $G$ of a set of feature maps $F_1, \cdots, F_n$ such that it has elements
\begin{equation}
	G_{ij} = \mbox{mean}(F_i \cdot F_j),
\end{equation}
where the product $F_i \cdot F_j$ is element-wise. $T_G$ is defined as the concatenation of the flattened Gram matrices computed for the feature maps before each pooling operation in VGG19. Note that the size of the Gram matrices depends on the number of feature maps (channels), not their size; thus $T_G$ is independent of input image size.

\label{ssec:example_summary_func}

\paragraph*{Statistics and Fourier transforms of image bins.}
While the neural summary function performs quite well, we find that in some cases we can improve upon it.
A simple idea for a summary function is to use the (RGB) mean of the entire image; an improvement is to subdivide the image into $k$ bins (regions) and compute the mean of each region. We found concentric bins perform well for isotropic materials, and vertical bins are appropriate for anisotropic highlights (e.g. brushed metal). Furthermore, we can additionally use a fast Fourier transform of the entire image or within bins. Note that automatic computation of derivatives is possible with the FFT, and supported by the PyTorch framework. In our current results, we use a summary function that combines the means and 1D FFTs of 64 vertical bins for the brushed metal example; all other examples use the neural summary function combined with simple image mean.

	\section{Bayesian Inference of Material Parameters}
\label{sec:bayesian}
In what follows, we first describe a Bayesian formulation of the estimation problem in terms of a posterior distribution. Next, we discuss how to use the posterior for point estimation in a maximum a-posteriori (MAP) framework, and how the Markov-Chain Monte Carlo (MCMC) methods for Bayesian inference extend the point estimate approach by sampling from the posterior.

\subsection{Bayesian formulation}
\label{ssec:point_sec}
We treat the procedural model parameters $\btheta$ as random variables with corresponding probability distributions.

We first introduce a \emph{prior} probability distribution $p(\btheta)$ of the parameters, reflecting our pre-existing beliefs about the likelihood values of the unknown parameters. For example, for most material categories, we know what range the albedo color and roughness coefficients of the material should typically be in.

Further, we model the $\approx$ operator from Eq.~\eqref{eq:approx} as an error distribution. Specifically, we postulate that the difference between the simulated image summary $\summ(f(\btheta, \bz))$ and the target 
image summary $\summ(\target)$ follows a known probability distribution.
In practice, we use a (multi-variate) normal distribution with zero mean and the covariance $\bsigma_e$:
\begin{equation}
\summ(f(\btheta, \bz)) - \summ(\target) \sim \mathcal{N}(0, \bsigma_e).
\end{equation}
Our experiments indicate that this simple error distribution works well in practice, and we regard $\bsigma_e$ as a hyper-parameter and set it manually.

We also have multiple options in handling the random vector $\bz$. While it is certainly theoretically possible to estimate it, we are not really interested in its values;  we find it simpler and more efficient to simply choose $\bz$ randomly, fix it, and assume it known during the process of estimating the ``interesting'' parameters $\btheta$.

Under these assumptions, according to the Bayes theorem, we can write down the posterior probability of parameters $\btheta$, conditional on the known values of $\target$ and $\bz$, as:
\begin{equation} \label{eq:posterior}
p(\btheta | \target, \bz) \propto \mathcal{N}\left[\summ(f(\btheta, \bz)) - \summ(\target); 0, \bsigma_e\right] p(\btheta),
\end{equation}
where the right side does not need to be normalized; the constant factor has no effect on parameter estimates.
For numerical stability, we compute the negative log posterior, viz. $-\log p(\btheta | \target, \bz)$, in practice. Equation~\eqref{eq:posterior} also holds when $\btheta$ involves discrete parameters, as long as the prior is properly defined as a product of a continuous pdf $p(\btheta_c)$ and a discrete probability $p(\btheta_d)$.

\subsection{Point estimate of parameter values}

A point estimate of the parameter vector can be modeled in the maximum a-posteriori (MAP) framework. We simply estimate the desired parameter values $\btheta$ as the maximum of the posterior pdf $p(\btheta | \target, \bz)$ given by Eq.~\eqref{eq:posterior}. For continuous $\btheta$, this problem can be solved using standard non-linear optimization algorithms. In the presence of discrete parameters, there is no single accepted solution. While various heuristics could be used, our sampling approach described below provides a cleaner solution to discrete parameter estimation.

\begin{algorithm}[t]
	\caption{\label{alg:mcmc_sample}
		MCMC sampling of material parameters $(\btheta_c, \btheta_d)$
	}
	\SetAlgoLined
	\SetCommentSty{mycmtfn}
	\SetKwComment{tccinline}{// }{}
	\SetKwFunction{sample}{samplePosterior}
	\sample{$N$, $\pc$, $\btheta_c^{(1)}$, $\btheta_d^{(1)}$}{\\
		\KwIn{Sample count $N$, probability $\pc$ for sampling continuous parameters, initial continuous parameters~$\btheta_c^{(1)}$ and discrete ones~$\btheta_d^{(1)}$}
		\KwOut{$N$ material parameter estimates $\{ (\btheta_c^{(t)}, \btheta_d^{(t)}) \,:\, 1 \leq t \leq N \}$}
		\Begin{
			\For{$t = 1$ to $(N - 1)$}{
				Draw $\xi \sim U[0, 1)$\\
				\uIf(\tccinline*[f]{Mutate continuous parameters}){$\xi < \pc$}{
					$\btheta_c' \gets \mathrm{ContinuousSample}(\btheta_c^{(t)})$ \label{alg:line:malaSample}\\
					$\btheta_d' \gets \btheta_d^{(t)}$
				}
				\Else(\tccinline*[f]{Mutate discrete parameters}){
					$\btheta_c' \gets \btheta_c^{(t)}$\\
					$\btheta_d' \gets \mathrm{DiscreteSample}(\btheta_d^{(t)})$ \label{alg:line:mhSample}\\
				}
				$(\btheta_c^{(t + 1)}, \btheta_d^{(t + 1)}) \gets \mathrm{MetropolisHasting}((\btheta_c', \btheta_d'),\ (\btheta_c^{(t)}, \btheta_d^{(t)}))$
				\label{alg:line:mhAccRej}
			}
			\KwRet{$\{ (\btheta_c^{(1)}, \btheta_d^{(1)}), \ldots, (\btheta_c^{(N)}, \btheta_d^{(N)}) \}$}
		}
	}
\end{algorithm}

\subsection{Markov-Chain Monte Carlo Sampling of the Posterior}
\label{ssec:bayesian}
Although the point estimate approach gives satisfactory results in many cases, it is not without problems. For example, since a perfect match between a procedural material and a photograph is generally impossible, it can be desirable to have a set of imperfect matches for the user to choose from. Further, there could be an entire subset of the parameter space giving solutions of approximately equivalent fit under the target view and lighting; however, these may look quite different from each other in other configurations, and a user may want to explore those differences. Lastly, with the presence of discrete parameters, it is not obvious how to solve the maximization in Eq.~\eqref{eq:posterior} efficiently.

In this paper, we use the well-known technique of full Bayesian inference, sampling the posterior pdf defined in Eq.~\eqref{eq:posterior} using Markov-Chain Monte Carlo (MCMC) techniques, specifically Metropolis-Hasting (MH)~\cite{Hastings}, Hamiltonian Monte Carlo (HMC)~\cite{Betancourt2017}, and Metropolis-adjusted Langevin algorithm (MALA)~\cite{MALA}. While well explored in statistics and various scientific fields, to our knowledge, this technique has not been used for the inference of material parameters.

The goal of the sampling is to explore the posterior with many (typically thousands or more) samples, each of which represents a material parameter vector consistent with the target image. Plotting these samples projected into two dimensions (for a given pair of parameters) gives valuable insight into similarity structures. Furthermore, interactively clicking on samples and observing the predicted result can help a user to quickly zoom in on a preferred solution, which an automatic optimization algorithm is fundamentally incapable of.

Algorithm~\ref{alg:mcmc_sample} summarizes our MCMC sampling process. At each iteration, we mutate either the continuous parameters (with probability $\pc$) or the discrete ones (with probability $1 - \pc$).
For the former case, we utilize the gradient of the log pdf with respect to $\btheta_c$ to efficiently obtain a new proposal $\btheta_c'$ (Line~\ref{alg:line:malaSample}).
Our implementation uses MALA for this process, although HMC could also work.
For the latter case, we obtain a new proposal $\btheta_d'$ of the discrete parameters, currently by uniformly sampling their joint probability mass function (Line~\ref{alg:line:mhSample}).
Upon obtaining a full proposal, we use the standard Metropolis-Hasting rule (Line~\ref{alg:line:mhAccRej}) to stochastically select the new sample $(\btheta_c^{(t + 1)}, \btheta_d^{(t + 1)})$ by either accepting the newly proposed $(\btheta_c', \btheta_d')$ or (rejecting the proposal and) keeping the previous sample $(\btheta_c^{(t)}, \btheta_d^{(t)})$.

	\section{Material Models and Results}
\label{sec:results}
\begin{figure}[t]
	\centering
	\addtolength{\tabcolsep}{-3pt}
	\begin{tabular}{c}
		\includegraphics[width=0.98\columnwidth]{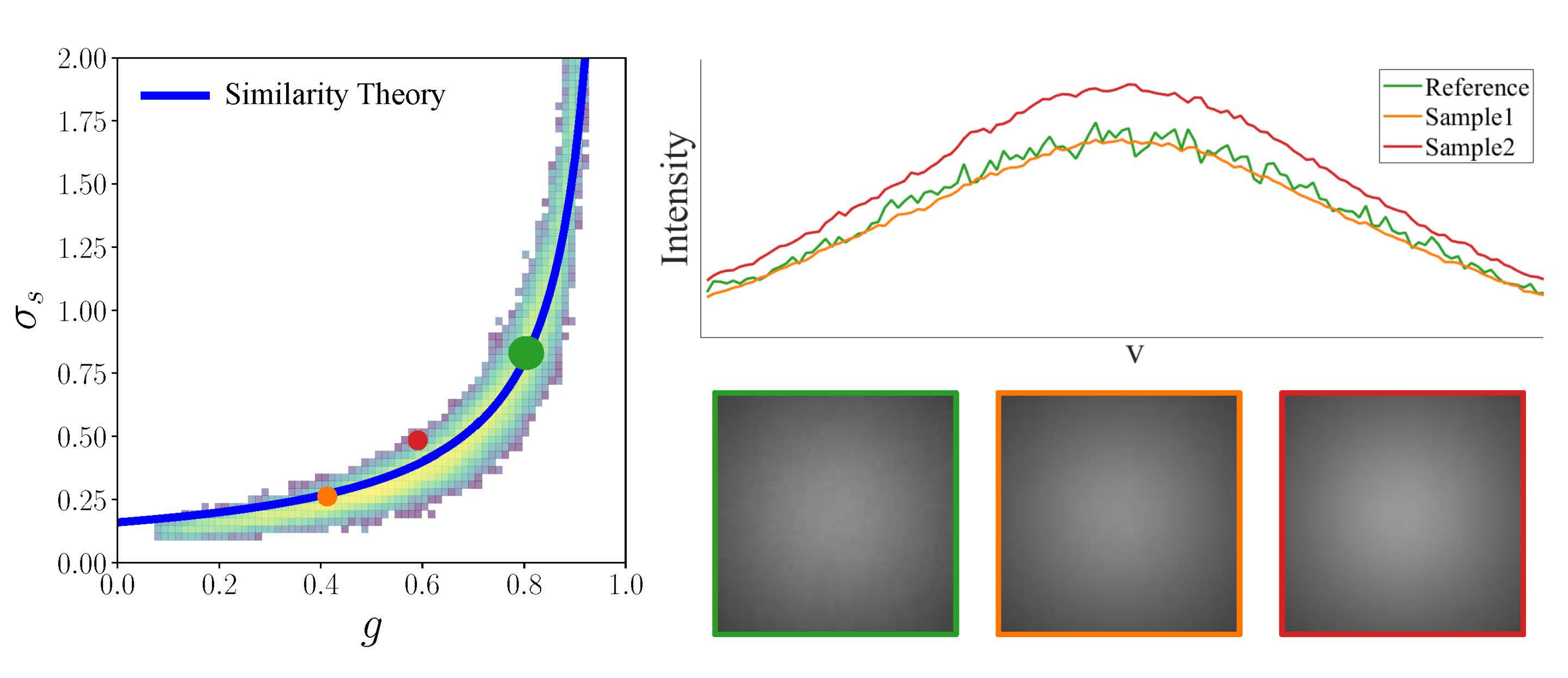}
	\end{tabular}
	\captionsetup{labelfont=bf,textfont=it}
	\caption{\label{fig:scatter}
		A motivating example of a scattering material with two estimated parameters (scattering coefficient and phase function parameter). The posterior distribution sampled with our method for three synthetic input images is able to detect the full structure of the parameter space, matching the predictions from similarity theory.
	}
\end{figure}


We now demonstrate the effectiveness of our technique by fitting several procedural material models to a mix of synthetic and real target images.

Our forward evaluation process uses collocated camera and light.
This configuration closely matches a mobile phone camera with flash (which is used for most of the real target images) and simplifies some BRDF formulations (because the incoming, outgoing, and half-way vectors are all identical).
Further, we assume that the distance between camera and sample is known as it is generally easy to measure or estimate.
The knowledge of the camera field of view allows us to compute the physical scale of the resulting pixels.
Lastly, we treat light intensity and camera vignetting (expressed as an image-space Gaussian function) as (unknown) parameters of the forward evaluation process so that they do not need to be calibrated.
Our parameter inference framework presented in \S\ref{sec:summary_func} and \S\ref{sec:bayesian} is not limited to this specific setup.

All the procedural material models we used, which will be detailed in \S\ref{ssec:proc_models}, are implemented using \textsf{PyTorch} which automatically provides GPU acceleration and computes derivatives through backpropagation. 
For all material parameter inference tasks, our forward evaluation generates $512 \times 512$ images.
Notice that the recovered parameters can then be used to generate results with much higher resolution because the procedural models are generally resolution-independent.

\subsection{Similarity Relations in Translucency}

As a motivating example, we first illustrate the behavior of the MCMC material parameter estimation process on the case of a homogeneous translucent material with two varying parameters.
In this example, the shape of the posterior can be analytically derived (using the similarity theory) and easily plotted. This serves as a demonstration and validation of our approach.

Specifically, the material parameter space of translucent materials under the radiative transfer framework~\cite{chandrasekhar1960radiative} is known to be approximately over-complete~\cite{Zhao:2014:HSR}.
Specifically, two sets of parameters $(\sigma_s, \sigma_a, g)$ and $(\sigma_s^*, \sigma_a^*, g^*)$ satisfying the following \emph{similarity relation} usually yield similar final appearances:
\begin{equation}
	\label{eq:similarity_rel}
	\sigma_a = \sigma_a^*, \quad (1 - g)\,\sigma_s = (1 - g^*)\,\sigma_s^*,
\end{equation}
where $\sigma_a$ and $\sigma_s$ are, respectively, the absorption and scattering coefficients, and $g$ is the first Legendre moment of the phase function.
We show in Figure~\ref{fig:scatter} that applying our Bayesian inference method to $\sigma_s$ and $g$ (with fixed $\sigma_a$) computes a posterior distribution that agrees well with the predicted similarity relation~\eqref{eq:similarity_rel}.

\begin{figure*}[t]
	\centering
	\addtolength{\tabcolsep}{-4.5pt}
	\begin{tabular}{ccccccccc}
		target & sample-1 & sample-2 & sample-3 & & target & sample-1 & sample-2 & sample-3
		\\
		\begin{overpic}[width=\resultwidth]{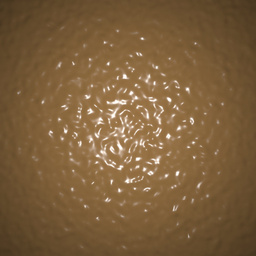}
			\imglabel{Bump-1}
		\end{overpic} &
		\includegraphics[width=\resultwidth]{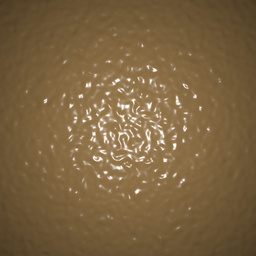} &
		\includegraphics[width=\resultwidth]{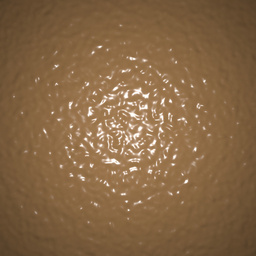} &
		\includegraphics[width=\resultwidth]{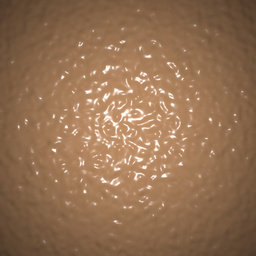} &
		&
		\begin{overpic}[width=\resultwidth]{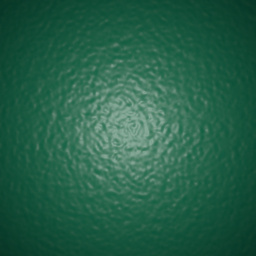}
			\imglabel{Bump-2}
		\end{overpic} &
		\includegraphics[width=\resultwidth]{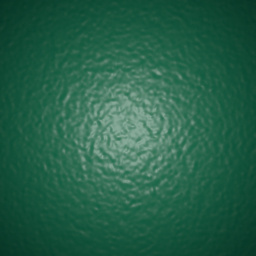} &
		\includegraphics[width=\resultwidth]{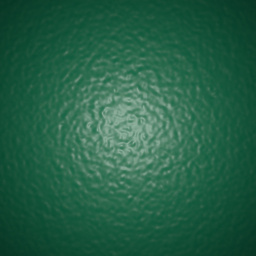} &
		\includegraphics[width=\resultwidth]{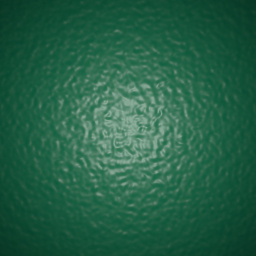}
		\\
		\begin{overpic}[width=\resultwidth]{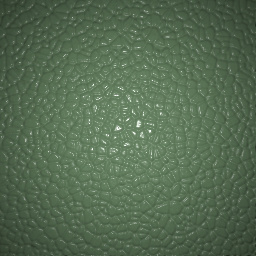}
			\imglabel{Leather-1}
		\end{overpic} &
		\includegraphics[width=\resultwidth]{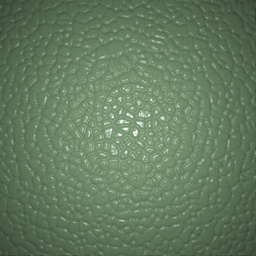} &
		\includegraphics[width=\resultwidth]{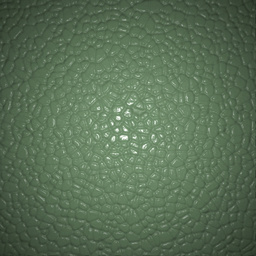} &
		\includegraphics[width=\resultwidth]{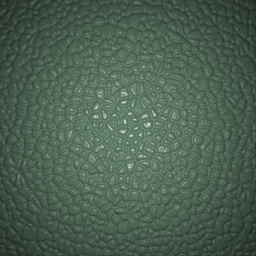} &
		&
		\begin{overpic}[width=\resultwidth]{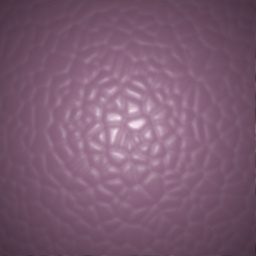}
			\imglabel{Leather-2}
		\end{overpic} &
		\includegraphics[width=\resultwidth]{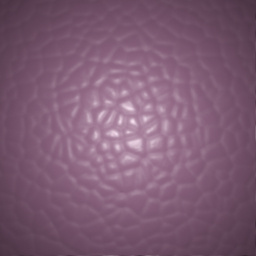} &
		\includegraphics[width=\resultwidth]{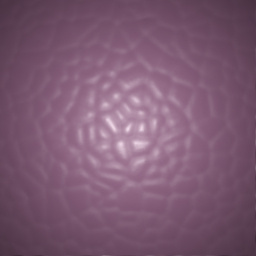} &
		\includegraphics[width=\resultwidth]{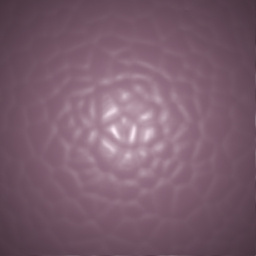}
		\\
		\begin{overpic}[width=\resultwidth]{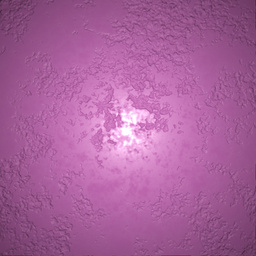}
			\imglabel{Plaster-1}
		\end{overpic} &
		\includegraphics[width=\resultwidth]{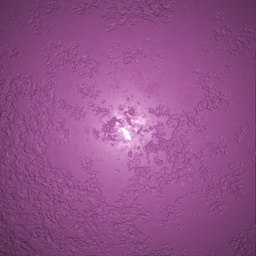} &
		\includegraphics[width=\resultwidth]{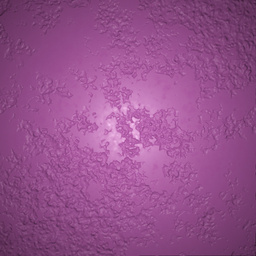} &
		\includegraphics[width=\resultwidth]{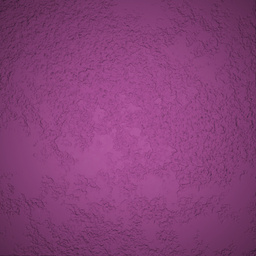} &
		&
		\begin{overpic}[width=\resultwidth]{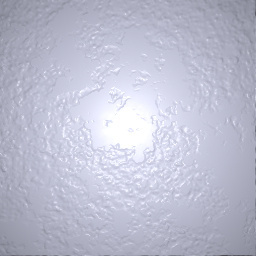}
			\imglabel{Plaster-2}
		\end{overpic} &
		\includegraphics[width=\resultwidth]{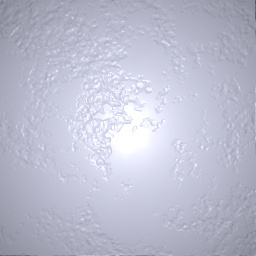} &
		\includegraphics[width=\resultwidth]{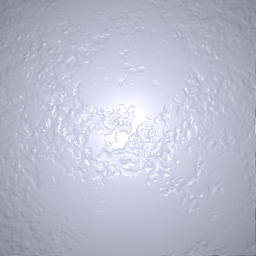} &
		\includegraphics[width=\resultwidth]{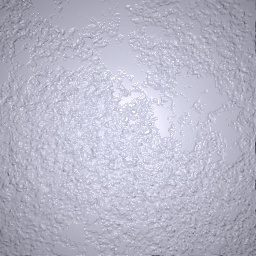}
		\\
		\begin{overpic}[width=\resultwidth]{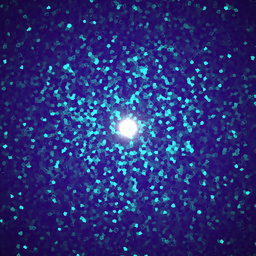}
			\imglabel{Metallicflake-1}
		\end{overpic} &
		\includegraphics[width=\resultwidth]{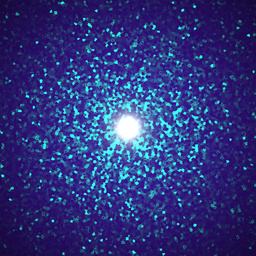} &
		\includegraphics[width=\resultwidth]{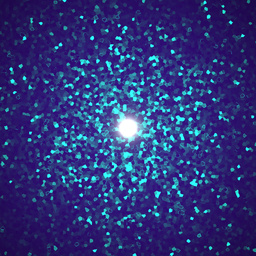} &
		\includegraphics[width=\resultwidth]{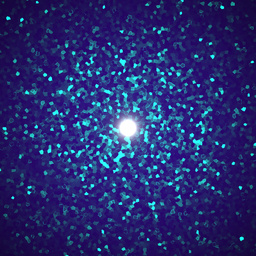} &
		&
		\begin{overpic}[width=\resultwidth]{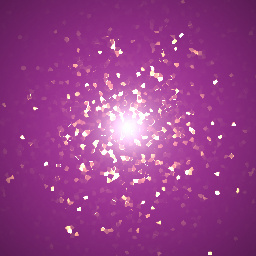}
			\imglabel{Metallicflake-2}
		\end{overpic} &
		\includegraphics[width=\resultwidth]{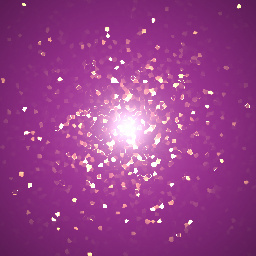} &
		\includegraphics[width=\resultwidth]{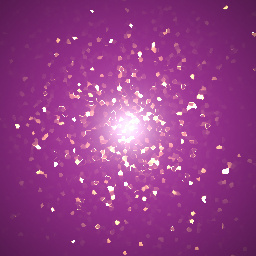} &
		\includegraphics[width=\resultwidth]{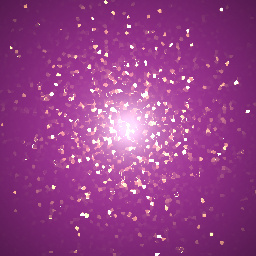}
		\\
		\begin{overpic}[width=\resultwidth]{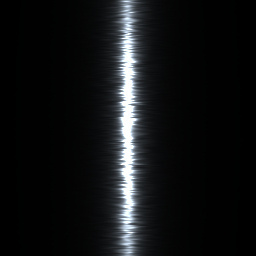}
			\imglabel{Brushmetal-1}
		\end{overpic} &
		\includegraphics[width=\resultwidth]{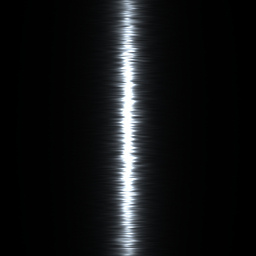} &
		\includegraphics[width=\resultwidth]{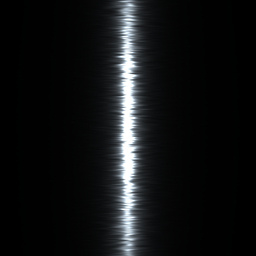} &
		\includegraphics[width=\resultwidth]{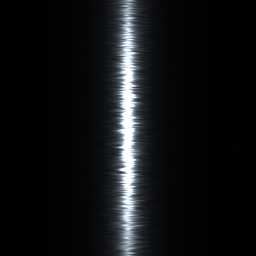} &
		&
		\begin{overpic}[width=\resultwidth]{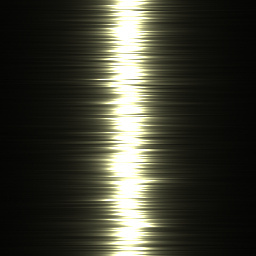}
			\imglabel{Brushmetal-2}
		\end{overpic} &
		\includegraphics[width=\resultwidth]{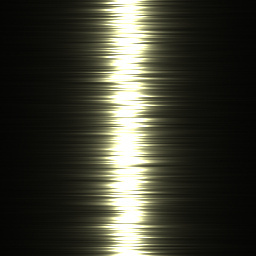} &
		\includegraphics[width=\resultwidth]{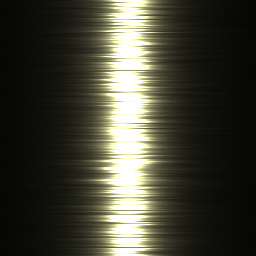} &
		\includegraphics[width=\resultwidth]{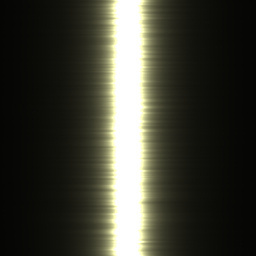}
		\\
		\begin{overpic}[width=\resultwidth]{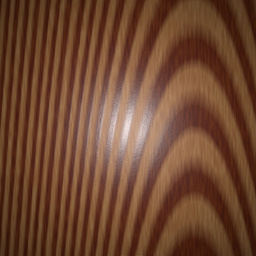}
			\imglabel{Wood-1}
		\end{overpic} &
		\includegraphics[width=\resultwidth]{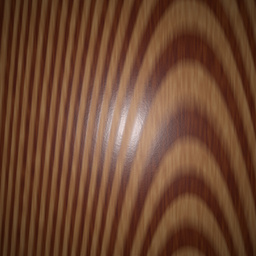} &
		\includegraphics[width=\resultwidth]{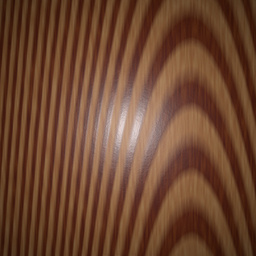} &
		\includegraphics[width=\resultwidth]{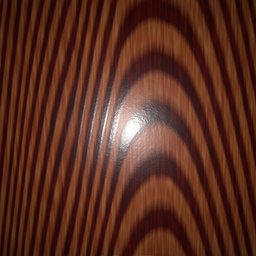} & &
		\begin{overpic}[width=\resultwidth]{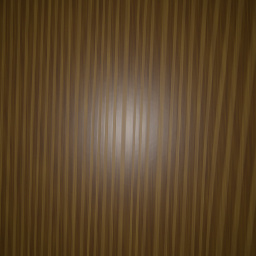}
			\imglabel{Wood-2}
		\end{overpic} &
		\includegraphics[width=\resultwidth]{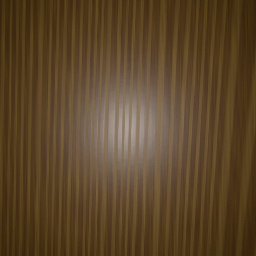} &
		\includegraphics[width=\resultwidth]{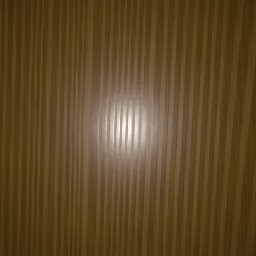} &
		\includegraphics[width=\resultwidth]{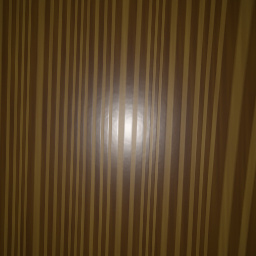}
	\end{tabular}
	\captionsetup{labelfont=bf,textfont=it}
	\caption{\label{fig:synth}
		\textbf{Results} of our MCMC sampling on \textbf{synthetic} inputs. Each row corresponds to two examples of a different material model. For each example, the first column is the synthetic target image. We show MCMC samples in the other columns, where sample-1 and sample-2 are chosen closer to the peak of the posterior distribution, and sample-3 is further away. More results please refer to supplemental materials.
	}
\end{figure*}

\begin{figure*}[t]
	\centering
	\addtolength{\tabcolsep}{-4.5pt}
	\begin{tabular}{cccccccc}
		target & sample-1 & sample-2 & sample-3 & target & sample-1 & sample-2 & sample-3
		\\
		\begin{overpic}[width=\resultwidth]{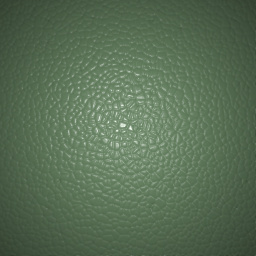}
			\imglabel{Leather-1}
		\end{overpic} &
		\includegraphics[width=\resultwidth]{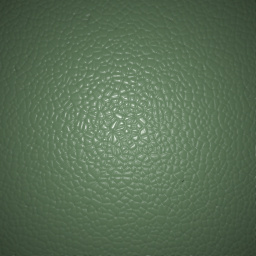} &
		\includegraphics[width=\resultwidth]{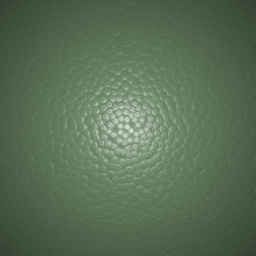} &
		\includegraphics[width=\resultwidth]{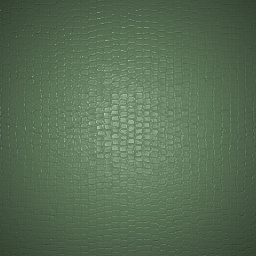} &
		\begin{overpic}[width=\resultwidth]{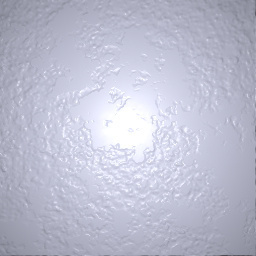}
			\imglabel{Plaster-2}
		\end{overpic} &
		\includegraphics[width=\resultwidth]{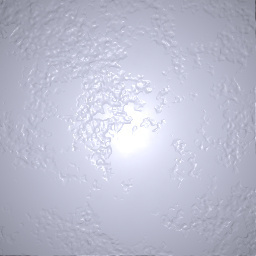} &
		\includegraphics[width=\resultwidth]{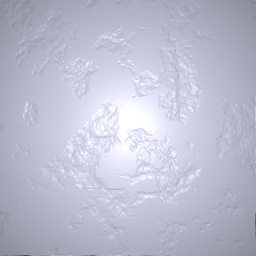} &
		\includegraphics[width=\resultwidth]{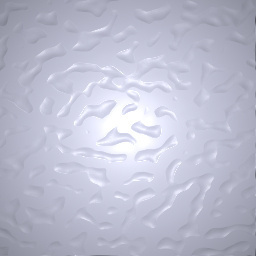}
		\\
		&
		\begin{overpic}[width=\resultwidth]{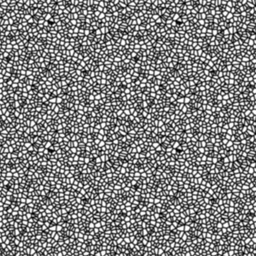}
			\put(0,0){\color{green}%
				\frame{\includegraphics[width=0.4\resultwidth]{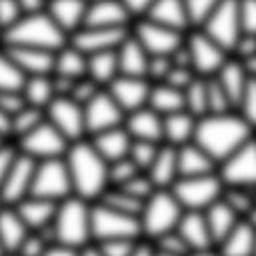}}}
		\end{overpic}
		&
		\begin{overpic}[width=\resultwidth]{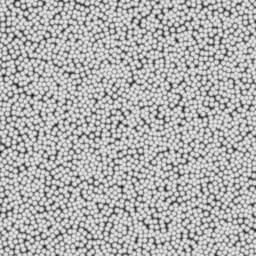}
			\put(0,0){\color{green}%
				\frame{\includegraphics[width=0.4\resultwidth]{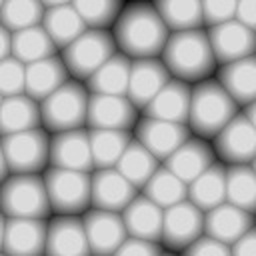}}}
		\end{overpic}
		&
		\begin{overpic}[width=\resultwidth]{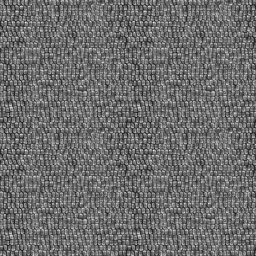}
			\put(0,0){\color{green}%
				\frame{\includegraphics[width=0.4\resultwidth]{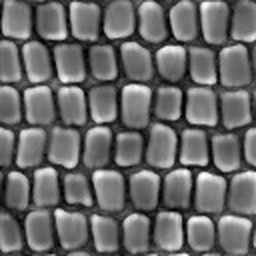}}}
		\end{overpic}
		&
		&
		\begin{overpic}[width=\resultwidth]{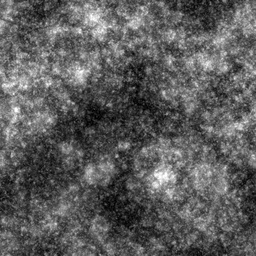}
			\put(0,0){\color{green}%
				\frame{\includegraphics[width=0.4\resultwidth]{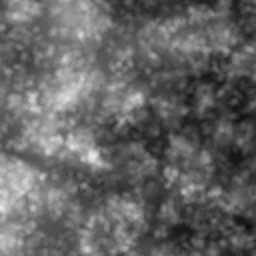}}}
		\end{overpic}
		&
		\begin{overpic}[width=\resultwidth]{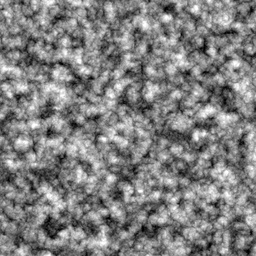}
			\put(0,0){\color{green}%
				\frame{\includegraphics[width=0.4\resultwidth]{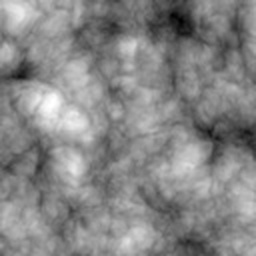}}}
		\end{overpic}
		&
		\begin{overpic}[width=\resultwidth]{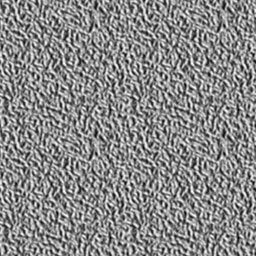}
			\put(0,0){\color{green}%
				\frame{\includegraphics[width=0.4\resultwidth]{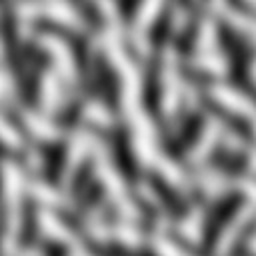}}}
		\end{overpic}
	\end{tabular}
	\captionsetup{labelfont=bf,textfont=it}
	\caption{\label{fig:discrete}
		\textbf{MCMC sampling with discrete parameters.} In these examples, we illustrate the ability of our sampling to handle discrete parameters. In both examples, one noise inputs used in the procedural model can be switched between several different types of noise. Out of the thousands of sampled solutions, we pick three that have different settings of the discrete parameter where the (log) pdf values decrease from sample-1 to sample-3.
	}
\end{figure*}

\begin{figure*}[t]
	\centering
	\addtolength{\tabcolsep}{-4.5pt}
	\begin{tabular}{ccccccccc}
		Photo & Sample-1 & Sample-2 & Sample-3 & & Photo & Sample-1 & Sample-2 & Sample-3
		\\
		\begin{overpic}[width=\resultwidth]{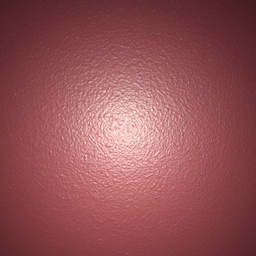}
			\imglabel{Bump-3}
		\end{overpic} &
		\includegraphics[width=\resultwidth]{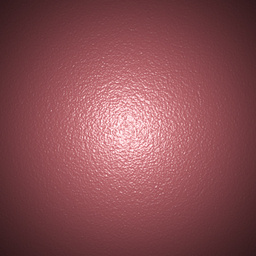} &
		\includegraphics[width=\resultwidth]{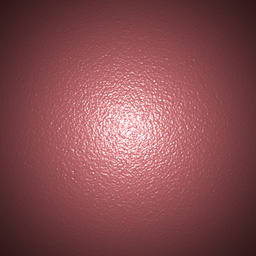} &
		\includegraphics[width=\resultwidth]{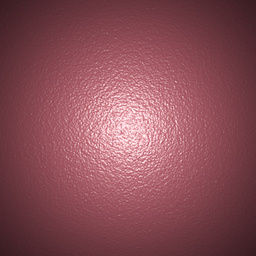} &
		&
		\begin{overpic}[width=\resultwidth]{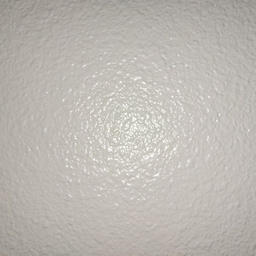}
			\imglabel{Bump-4}
		\end{overpic} &
		\includegraphics[width=\resultwidth]{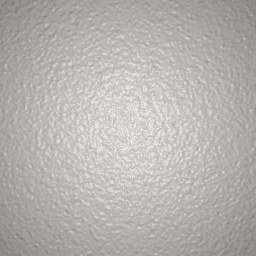} &
		\includegraphics[width=\resultwidth]{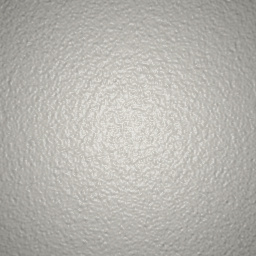} &
		\includegraphics[width=\resultwidth]{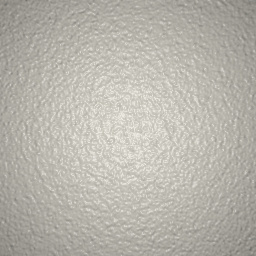}
		\\
		\begin{overpic}[width=\resultwidth]{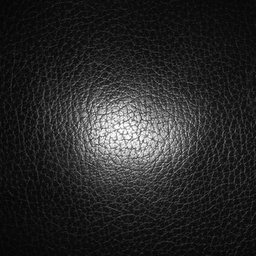}
			\imglabel{Leather-3}
		\end{overpic} &
		\includegraphics[width=\resultwidth]{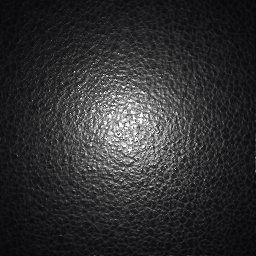} &
		\includegraphics[width=\resultwidth]{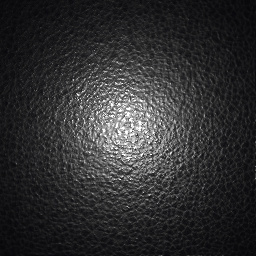} &
		\includegraphics[width=\resultwidth]{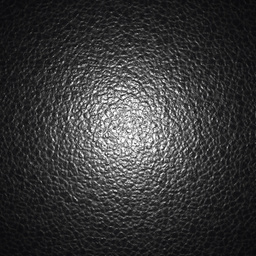} &
		&
		\begin{overpic}[width=\resultwidth]{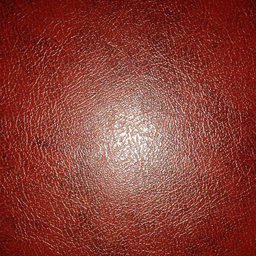}
			\imglabel{Leather-4}
		\end{overpic} &
		\includegraphics[width=\resultwidth]{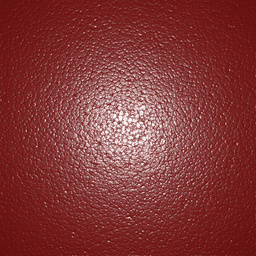} &
		\includegraphics[width=\resultwidth]{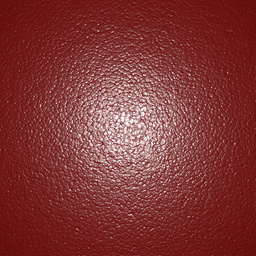} &
		\includegraphics[width=\resultwidth]{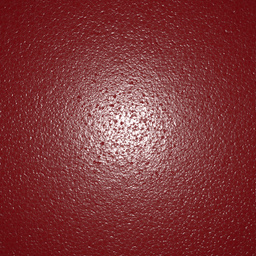}
		\\
		\begin{overpic}[width=\resultwidth]{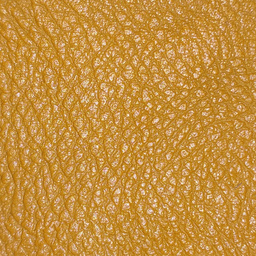}
			\imglabel{Leather-5}
		\end{overpic} &
		\includegraphics[width=\resultwidth]{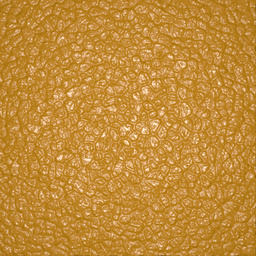} &
		\includegraphics[width=\resultwidth]{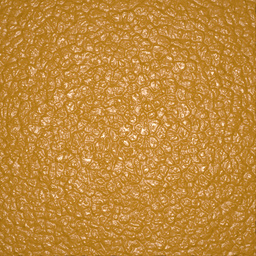} &
		\includegraphics[width=\resultwidth]{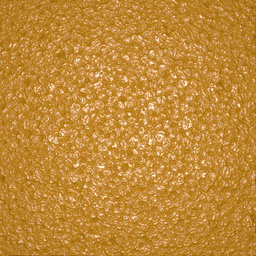} &
		&
		\begin{overpic}[width=\resultwidth]{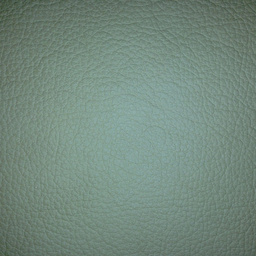}
			\imglabel{Leather-6}
		\end{overpic} &
		\includegraphics[width=\resultwidth]{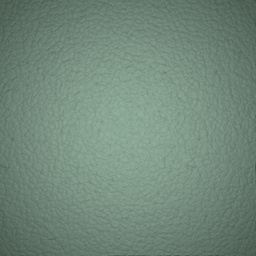} &
		\includegraphics[width=\resultwidth]{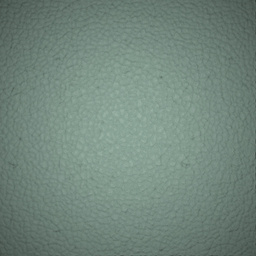} &
		\includegraphics[width=\resultwidth]{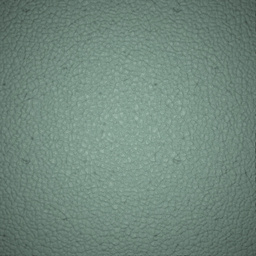}
		\\
		\begin{overpic}[width=\resultwidth]{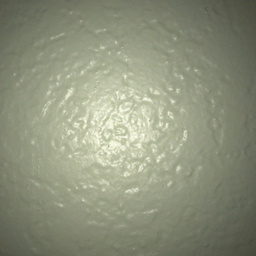}
			\imglabel{Plaster-3}
		\end{overpic} &
		\includegraphics[width=\resultwidth]{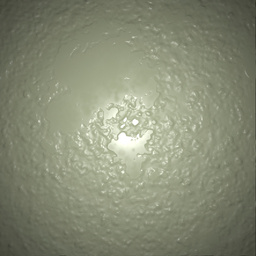} &
		\includegraphics[width=\resultwidth]{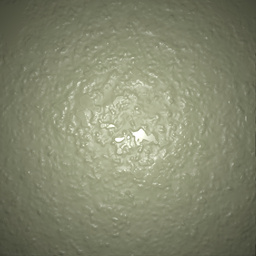} &
		\includegraphics[width=\resultwidth]{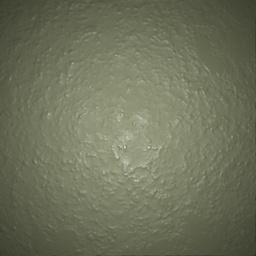} &
		&
		\begin{overpic}[width=\resultwidth]{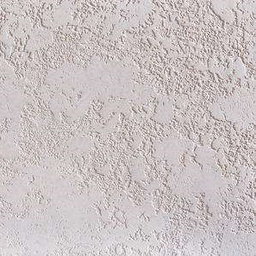}
			\imglabel{Plaster-4}
		\end{overpic} &
		\includegraphics[width=\resultwidth]{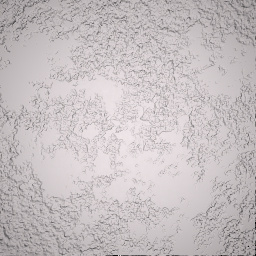} &
		\includegraphics[width=\resultwidth]{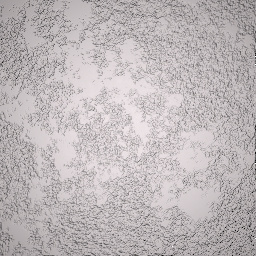} &
		\includegraphics[width=\resultwidth]{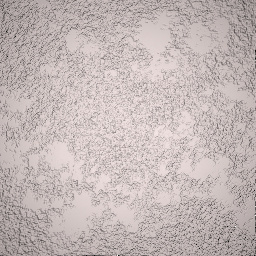}
		\\
		\begin{overpic}[width=\resultwidth]{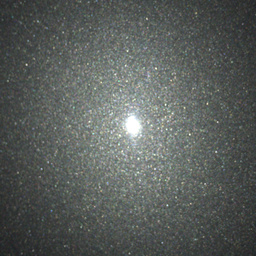}
			\imglabel{Metallicflake-3}
		\end{overpic} &
		\includegraphics[width=\resultwidth]{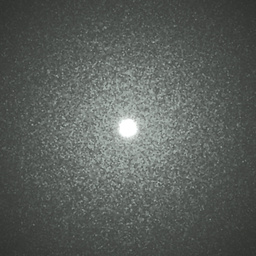} &
		\includegraphics[width=\resultwidth]{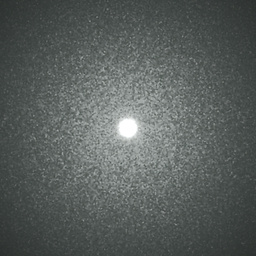} &
		\includegraphics[width=\resultwidth]{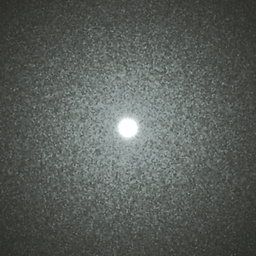} &
		&
		\begin{overpic}[width=\resultwidth]{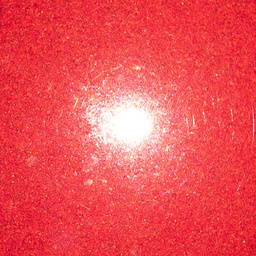}
			\imglabel{Metallicflake-4}
		\end{overpic} &
		\includegraphics[width=\resultwidth]{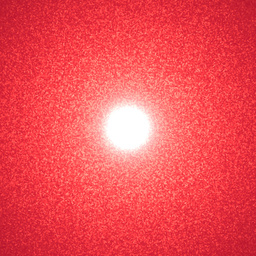} &
		\includegraphics[width=\resultwidth]{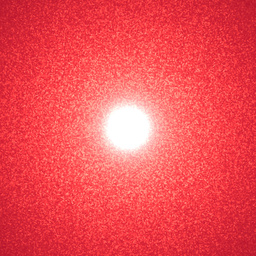} &
		\includegraphics[width=\resultwidth]{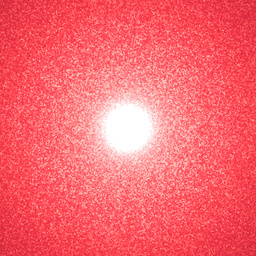}
		\\
		\begin{overpic}[width=\resultwidth]{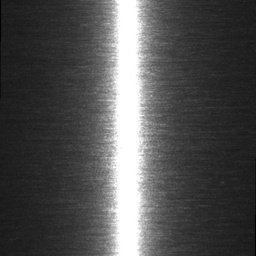}
			\imglabel{Brushmetal-3}
		\end{overpic} &
		\includegraphics[width=\resultwidth]{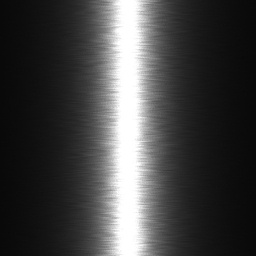} &
		\includegraphics[width=\resultwidth]{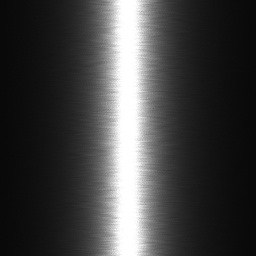} &
		\includegraphics[width=\resultwidth]{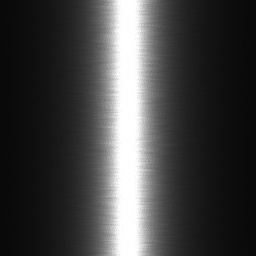} &
		&
		\begin{overpic}[width=\resultwidth]{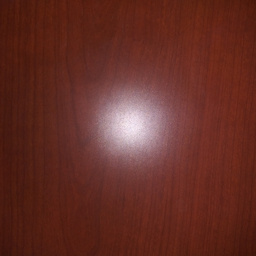}
			\imglabel{Wood-3}
		\end{overpic} &
		\includegraphics[width=\resultwidth]{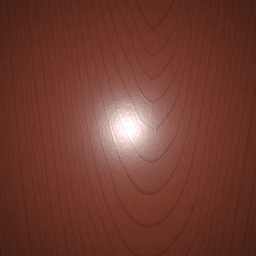} &
		\includegraphics[width=\resultwidth]{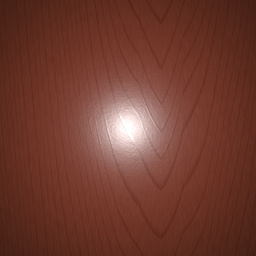} &
		\includegraphics[width=\resultwidth]{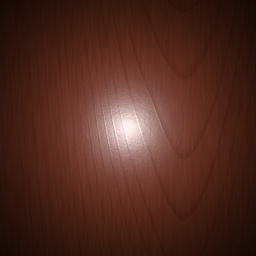}
		\\
		\begin{overpic}[width=\resultwidth]{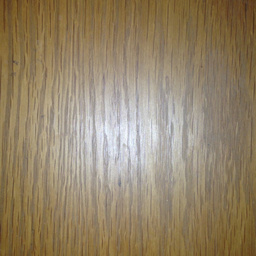}
			\imglabel{Wood-4}
		\end{overpic} &
		\includegraphics[width=\resultwidth]{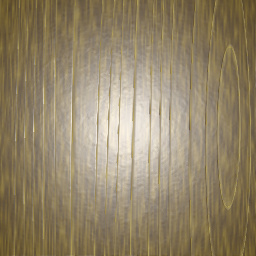} &
		\includegraphics[width=\resultwidth]{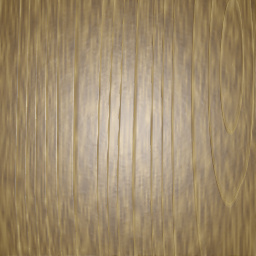} &
		\includegraphics[width=\resultwidth]{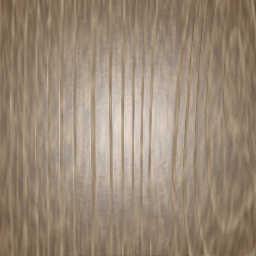} &
		&
		\begin{overpic}[width=\resultwidth]{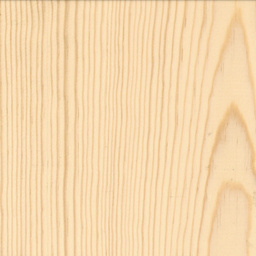}
			\imglabel{Wood-5}
		\end{overpic} &
		\includegraphics[width=\resultwidth]{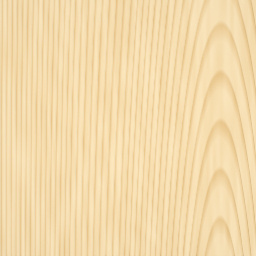} &
		\includegraphics[width=\resultwidth]{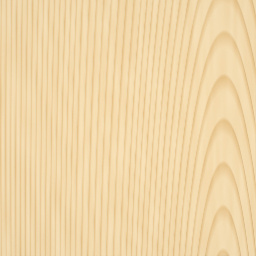} &
		\includegraphics[width=\resultwidth]{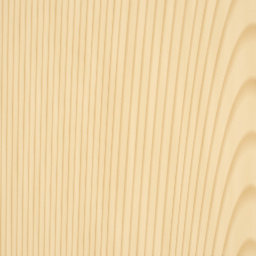}
	\end{tabular}
	\captionsetup{labelfont=bf,textfont=it}
	\caption{\label{fig:real}
		\textbf{Results} of our MCMC sampling on \textbf{real} inputs. For each example, the first column is the real target image (photo). We show MCMC samples in the other columns, where sample-1 and sample-2 are chosen closer to the peak of the posterior distribution, and sample-3 is further away. Note that the target images for Plaster-4 and Wood-5 are captured under natural illumination, while the corresponding synthetic images still assume collocated flash illumination; despite this mismatch, the estimated material parameters are still reasonable. Note, target images for Leather-4, Leather-6 and Wood-4 are from the publicly released dataset of \cite{Aittala2016}. For more results please refer to supplemental materials.
	}
\end{figure*}

\begin{table}[t]
	\centering
	\caption{\label{fig:performance}
		Performance statistics for our MCMC-based posterior sampling.
		The numbers are collected using a workstation equipped with an Intel i7-6800K six-core CPU and an Nvidia GTX 1080 GPU.  
	}
	\addtolength{\tabcolsep}{-3pt}
	\begin{tabular}{c|cccccc}
		\textbf{Material} & bump & leather & plaster & flakes & metal & wood\\
		\hline
		\textbf{\# params.} & 8 & 12 & 11 & 13 & 10 & 23\\
		\textbf{MCMC} (1k iter.) & 180s & 194s & 190s & 187s & 182s & 290s
	\end{tabular}
\end{table}


\subsection{Procedural Material Models}
\label{ssec:proc_models}
We show results generated using synthetic images in Figures~\ref{fig:synth} and \ref{fig:discrete} as well as real photographs (taken with different cameras) in Figure~\ref{fig:real}.
Please see the supplemental material for more results, including animations illustrating point estimations and sampling. Below we describe the six procedural models tested. Please refer to the supplement for additional detail and a \textsf{PyTorch} implementation. For each parameter, we define a reasonable truncated Gaussian distribution as its prior (also see supplement). In most cases, the MCMC sampling starts from the peak of the prior. In some examples (e.g wood), we first run posterior maximization and then switch to sampling from the optimized point. We drop some number (typically 200 to 1000) of initial MCMC samples due to burn-in.

\paragraph*{Bumpy microfacet surface.}
This model depicts an opaque dielectric surface with an isotropic noise heightfield. We use a standard microfacet BRDF with the GGX normal distribution~\cite{Walter2007} combined with a normal map computed from an explicitly constructed heightfield. We assume that the Fresnel reflectance at normal incidence can be computed from a known index of refraction (a value of 1.5 is a good estimate for plastics). We assume an unknown roughness $r$ (GGX parameter $\alpha=r^2$) and a Lambertian diffuse term with unknown albedo $\rho$. This model is identical to Wang et al.~\cite{Wang2011}, except using the GGX instead of Beckmann microfacet distribution. The main practical difference from the capture setup in that paper is that we use a point light, instead of step-edge illumination.

The bumpy heightfield is constructed using an inverse Fourier process including: (i)~choosing a power spectrum in the continuous Fourier domain; (ii)~discretizing it onto a grid of complex numbers; (iii)~randomly choosing the phase of each texel on the grid (while keeping the chosen amplitude); and (iv)~applying an inverse fast Fourier transform whose
real component becomes the resulting heightfield.
At render time, we use the normal map derived from this heightfield.

\paragraph*{Leather and plaster.}
These materials can be modeled similarly as the aforementioned bumpy surfaces except for the computation of the heightfield and roughness.
For plaster, a fractal noise texture is scaled (in space and intensity) and thresholded (controlled by additional parameters) to produce both the heightfield and a roughness variation texture. For leather, on the contrary, a Voronoi cell map is used to get the effect of leather-like cells (with parameters for scaling and thresholding), and additional small-scale fractal noise is added.
Further, we use multiple (pre-generated) noise textures and Voronoi cell maps to diversify the micro-scale appearances that our models can produce.
The choice of these textures and maps is captured using a discrete parameter.
In Figure~\ref{fig:discrete}, we show a few example samples drawn from the posterior distributions using Algorithm~\ref{alg:mcmc_sample}.

\paragraph*{Brushed metal.} The brushed metal material extends the above bumpy surface, by introducing anisotropy to both the GGX normal distribution and the noise heightfield used to compute the normal map, while dropping the diffuse term. We make both the BRDF and the Fourier-domain Gaussian power spectrum anisotropic. The parameters of the model thus include two roughnesses, as well as two Fourier-domain standard deviations.  We make the anisotropic highlight vertical and centered in the target image.

\paragraph*{Metallic flakes.} Metallic paint with flakes is a stochastic material with multiple BRDF lobes (caused by light reflecting off the flakes). Our model involves three components, each being an isotropic microfacet lobe, to describe top coating, flakes and glow, respectively. The top coating is usually highly specular, and we make its roughness a model parameter. We assume an index of refraction of 1.5, implying a Fresnel (Schlick) reflectivity at normal incidence of 0.04. The flakes are chosen as Voronoi cells of a random blue-noise point distribution; they have a roughness parameter and varying normals chosen from the Beckmann distribution with an unknown roughness, and with unknown Fresnel reflectivity. The scale of the cell map is itself a (differentiable) parameter. Lastly, the glow is a component approximating the internal scattering between the top interface and the flakes, and has its own roughness, Fresnel reflectivity and a flat normal. An extra weight parameter linearly combines the flakes and the glow.

\paragraph*{Wood.} We \revision{also} created a partial \textsf{PyTorch} implementation of the comprehensive 3D wood model of Liu et al.~\cite{Liu2016}. This material is a 3D model of the growth rings of a tree, with a number of parameters controlling colors and widths of growth rings, as well as global distortions and small-scale noise features. The 3D wood is finally projected by a cutting plane to image space, defining diffuse albedo, roughness and height.

\setlength{\fboxrule}{2pt}
\newcommand\fboxg{\fcolorbox{green}{white}}
\newcommand\fboxr{\fcolorbox{red}{white}}

\begin{figure}[t]
	\centering
	\addtolength{\tabcolsep}{-4.5pt}
	\begin{tabular}{cccc}
		Photo & \textit{Leather} Prior & \textit{Plaster} Prior & \textit{Wood} Prior
		\\
		\begin{overpic}[width=0.85\resultwidth]{fig7/2_leather_5/target.jpg}
			\imglabel{Leather-5}
		\end{overpic} &
		\fboxg{\includegraphics[width=0.85\resultwidth]{fig7/2_leather_5/good1.jpg}} &
		\fboxr{\includegraphics[width=0.85\resultwidth]{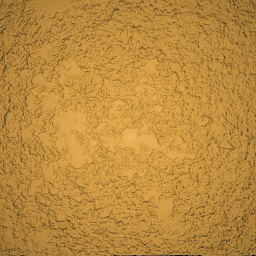}} &
		\fboxr{\includegraphics[width=0.85\resultwidth]{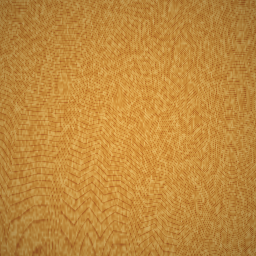}} 
		\\[5pt]
		\begin{overpic}[width=0.85\resultwidth]{fig7/6_wood_5/target.jpg}
			\imglabel{Wood-5}
		\end{overpic} &
		\fboxr{\includegraphics[width=0.85\resultwidth]{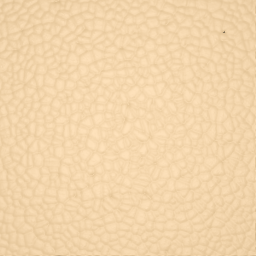}} &
		\fboxr{\includegraphics[width=0.85\resultwidth]{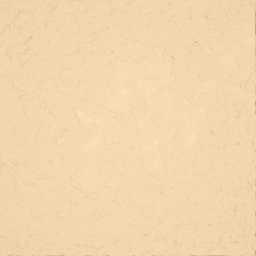}} &
		\fboxg{\includegraphics[width=0.85\resultwidth]{fig7/6_wood_5/good1.jpg}} 
	\end{tabular}
	\captionsetup{labelfont=bf,textfont=it}
	\caption{\label{fig:Mismatch}
		\revision{
			\textbf{Comparison} with mismatched forward models. With an inappropriate model as the prior, it would only match the global color but missing all the details.   
		}
	}
\end{figure}

\paragraph*{Mismatched models.}
Lastly, we demonstrate in Figure \ref{fig:Mismatch} the impact of forward procedural models.
Since these model-generating procedures are essentially material-specific priors, using mismatched models generally leads to results that match overall image statistics but with incorrect patterns.

\subsection{Additional Comparisons}

\begin{figure}[t]
	\centering
	\addtolength{\tabcolsep}{-4.5pt}
	\begin{tabular}{cccc}
		Photo & Ours & \cite{Aittala2016} & \cite{Aittala2016}-Maps
		\\
		\begin{overpic}[width=0.95\resultwidth]{fig7/2_leather_4/target.jpg}
			\imglabel{Leather-4}
		\end{overpic} &
		\includegraphics[width=0.95\resultwidth]{fig7/2_leather_4/good1.jpg} &
		\includegraphics[width=0.95\resultwidth]{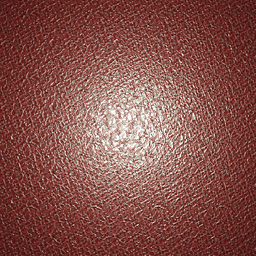} &
		\includegraphics[width=0.95\resultwidth]{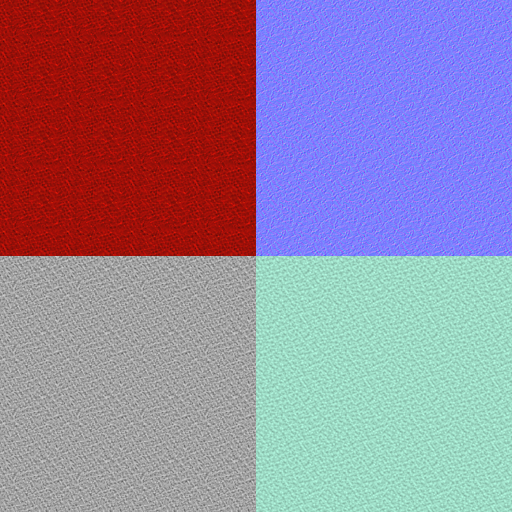}
		\\
		\begin{overpic}[width=0.95\resultwidth]{fig7/2_leather_6/target.jpg}
			\imglabel{Leather-6}
		\end{overpic} &
		\includegraphics[width=0.95\resultwidth]{fig7/2_leather_6/good1.jpg} &
		\includegraphics[width=0.95\resultwidth]{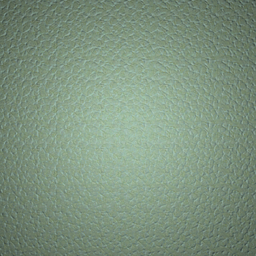} &
		\includegraphics[width=0.95\resultwidth]{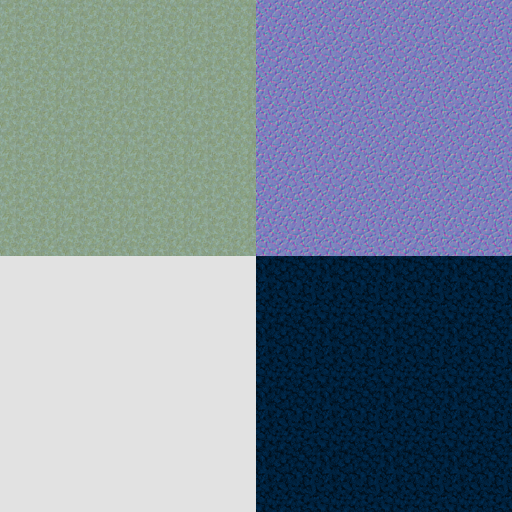}
		\\
		\begin{overpic}[width=0.95\resultwidth]{fig7/6_wood_4/target.jpg}
			\imglabel{Wood-4}
		\end{overpic} &
		\includegraphics[width=0.95\resultwidth]{fig7/6_wood_4/good1.jpg} &
		\includegraphics[width=0.95\resultwidth]{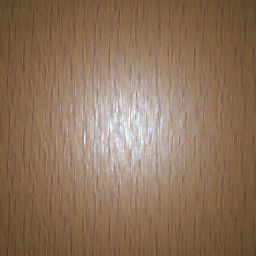} &
		\includegraphics[width=0.95\resultwidth]{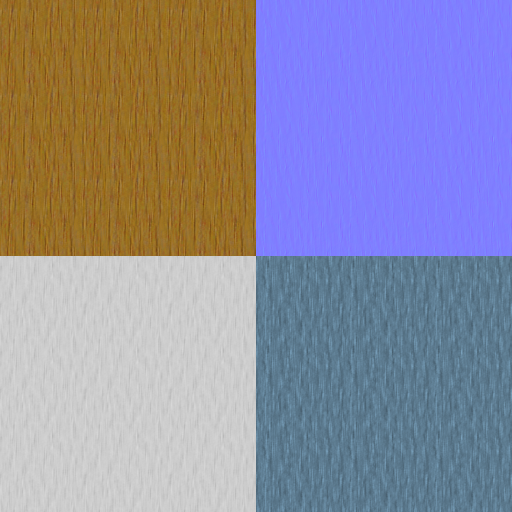}
	\end{tabular}
	\captionsetup{labelfont=bf,textfont=it}
	\caption{\label{fig:Aittala}
		\revision{
			\textbf{Comparison} with the Aittala et al.~\cite{Aittala2016}.
			Results in the third column are rendered using tiled versions of texture maps shown in the fourth column.
		}
	}
\end{figure}

\paragraph*{Comparison to Aittala et al.}
We first compare our technique to with one introduced by Aittala~et~al.~\cite{Aittala2016} in Figure \ref{fig:Aittala} using input photos published as supplemental materials from their work.
Their work uses the same VGG-based loss (summary function), but optimizes directly in texture space. Both methods manage to reproduce the overall pattern and reflectance of the input photos.
Thanks to the underlying procedural models, our method is able to synthesize larger results without visually obvious periodic patterns, and with more plausible global variation.

\begin{figure*}[t]
	\centering
	\addtolength{\tabcolsep}{-4.5pt}
	\begin{tabular}{ccccccccc}
		\raisebox{20pt}{\rotatebox{90}{\small Photo}}
		&
		\begin{overpic}[width=\resultwidth]{fig7/1_bump_3/target.jpg}
			\imglabeltop{Bump-3}
		\end{overpic}
		&
		\begin{overpic}[width=\resultwidth]{fig7/2_leather_3/target.jpg}
			\imglabeltop{Leather-3}
		\end{overpic}
		&
		\begin{overpic}[width=\resultwidth]{fig7/2_leather_6/target.jpg}
			\imglabeltop{Leather-6}
		\end{overpic}
		&
		\begin{overpic}[width=\resultwidth]{fig7/3_plaster_3/target.jpg}
			\imglabeltop{Plaster-3}
		\end{overpic}
		&
		\begin{overpic}[width=\resultwidth]{fig7/4_flake_4/target.jpg}
			\imglabeltop{Metallicflake-4}
		\end{overpic}
		&
		\begin{overpic}[width=\resultwidth]{fig7/5_metal_3/target.jpg}
			\imglabeltop{Brushmetal-3}
		\end{overpic}
		&
		\begin{overpic}[width=\resultwidth]{fig7/6_wood_3/target.jpg}
			\imglabeltop{Wood-3}
		\end{overpic}
		&
		\begin{overpic}[width=\resultwidth]{fig7/6_wood_4/target.jpg}
			\imglabeltop{Wood-4}
		\end{overpic}
		\\
		\raisebox{20pt}{\rotatebox{90}{\small Ours}} &
		\includegraphics[width=\resultwidth]{fig7/1_bump_3/good1.jpg} &
		\includegraphics[width=\resultwidth]{fig7/2_leather_3/good1.jpg} &
		\includegraphics[width=\resultwidth]{fig7/2_leather_6/good1.jpg} &
		\includegraphics[width=\resultwidth]{fig7/3_plaster_3/good1.jpg} &
		\includegraphics[width=\resultwidth]{fig7/4_flake_4/good1.jpg} &
		\includegraphics[width=\resultwidth]{fig7/5_metal_3/good1.jpg} &
		\includegraphics[width=\resultwidth]{fig7/6_wood_3/good1.jpg} &
		\includegraphics[width=\resultwidth]{fig7/6_wood_4/good1.jpg}
		\\
		\raisebox{10pt}{\rotatebox{90}{\small \cite{Hu2019}}} &
		\includegraphics[width=\resultwidth]{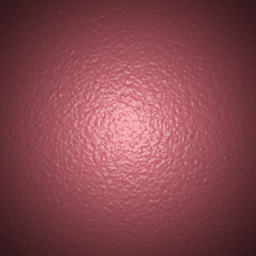} &
		\includegraphics[width=\resultwidth]{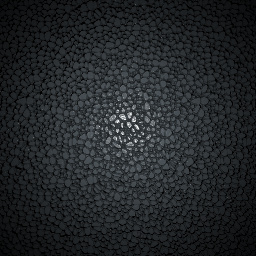} &
		\includegraphics[width=\resultwidth]{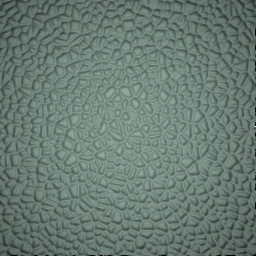} &
		\includegraphics[width=\resultwidth]{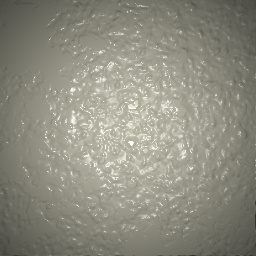} &
		\includegraphics[width=\resultwidth]{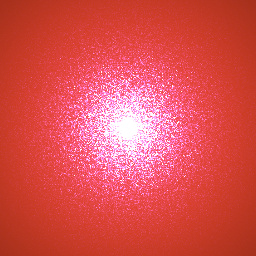} &
		\includegraphics[width=\resultwidth]{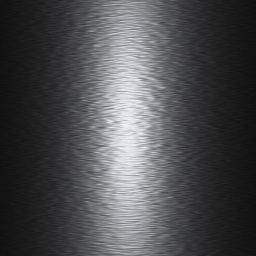} &
		\includegraphics[width=\resultwidth]{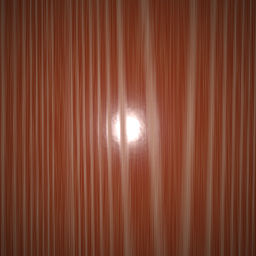} &
		\includegraphics[width=\resultwidth]{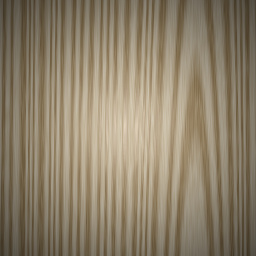}
	\end{tabular}
	\captionsetup{labelfont=bf,textfont=it}
	\caption{\label{fig:Hu}
		\textbf{Comparison} to the forward neural prediction method of Hu et al. \cite{Hu2019}, where we apply their network structure with our BRDFs and lighting conditions. The photo (top) is better matched by our MCMC sampling results (middle) than their prediction (bottom), which moreover tends to become worse for more complex BRDF models and with more parameters. On the other hand, \cite{Hu2019} can be used as an efficient initialization of our sampling, as shown in Figure \ref{fig:Hu2}.
	}
\end{figure*}

\begin{figure}[t]
	\centering
	\addtolength{\tabcolsep}{-3pt}
	\begin{tabular}{cccc}
		\raisebox{10pt}{\includegraphics[width=0.25\columnwidth]{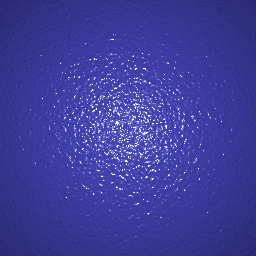}} &
		\raisebox{-4pt}{\includegraphics[width=0.32\columnwidth]{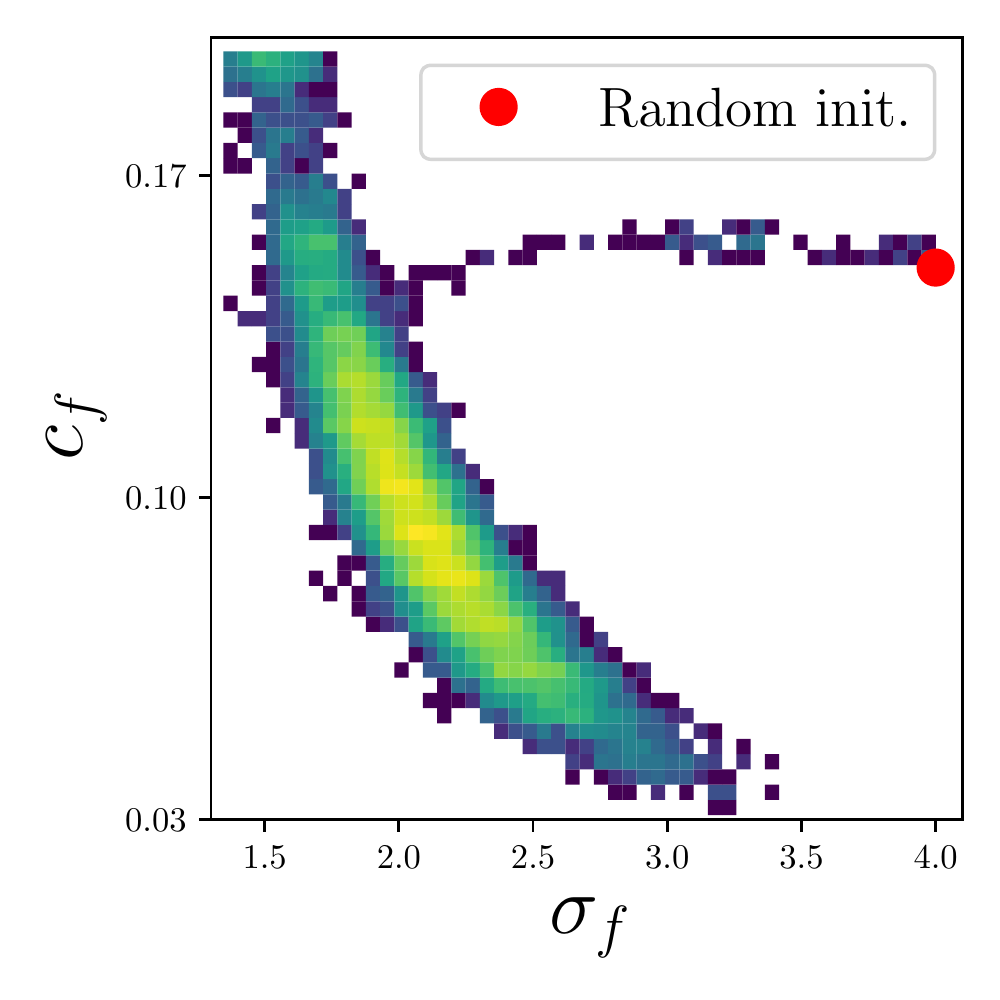}} &
		\raisebox{-4pt}{\includegraphics[width=0.32\columnwidth]{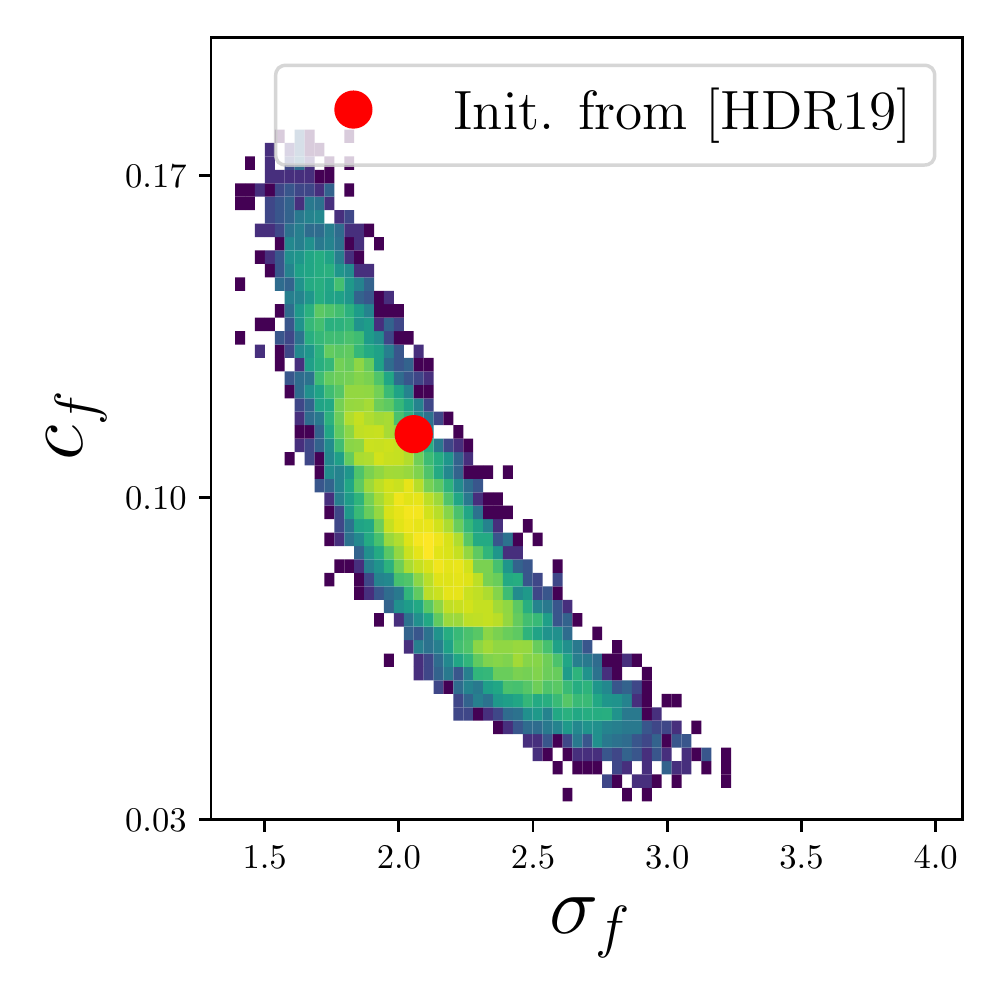}} &
		\raisebox{16pt}{\rotatebox{90}{$N=22500$}} \\
		Target & (a) & (b)
	\end{tabular}
	\captionsetup{labelfont=bf,textfont=it}
	\caption{\label{fig:Hu2}
		\textbf{Initialization} of our sampling with the method of Hu et al. \cite{Hu2019} on a synthetic bumpy surface example. The figure shows joint posterior distributions over two parameters using different initializations: a random initialization drawn from the prior (a) and the prediction of \cite{Hu2019} (b). As we can see, starting from the result of \cite{Hu2019} can shorten the burn-in phase of the MCMC sampling process.
	}
\end{figure}

\begin{figure}[t]
	\centering
	\addtolength{\tabcolsep}{-3pt}
	\begin{tabular}{c}
		\includegraphics[width=0.9\columnwidth]{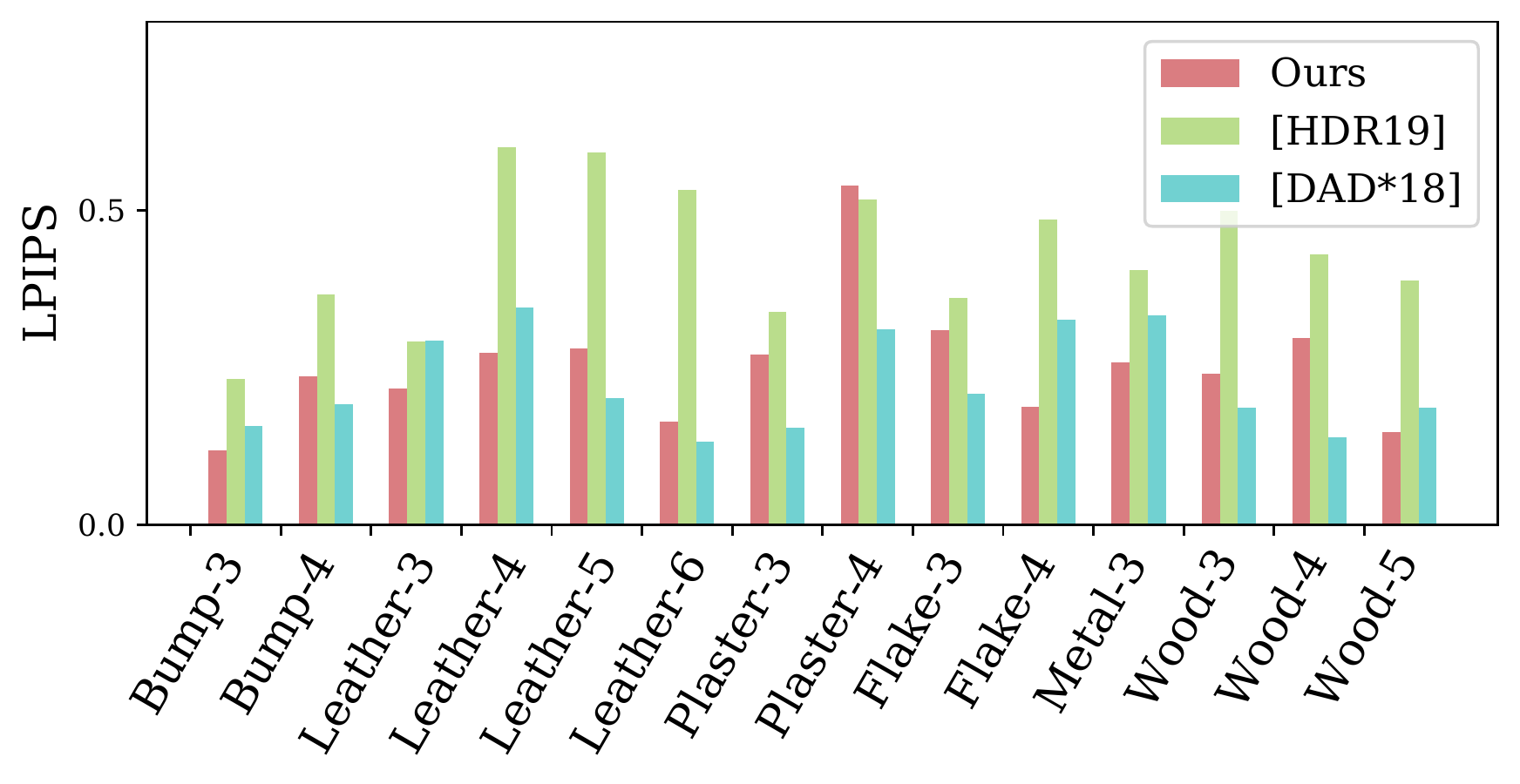}
	\end{tabular}
	\captionsetup{labelfont=bf,textfont=it}
	\caption{\label{fig:Comp}
		\revision{
			\textbf{Quantitative evaluation.} The LPIPS of our results are consistently better than \cite{Hu2019}. Some LPIPS values from \cite{Deschaintre2018} are better than ours, since (as per-pixel methods) they can better match the noise patterns in the textures.
		}
	}
\end{figure}
\begin{figure*}[t]
	\centering
	\addtolength{\tabcolsep}{-4.5pt}
	\begin{tabular}{ccccccccc}
		Photo & Ours & \cite{Deschaintre2018} & \cite{Deschaintre2018}-Maps & & Photo & Ours & \cite{Deschaintre2018} & \cite{Deschaintre2018}-Maps
		\\
		\begin{overpic}[width=\resultwidth]{fig7/1_bump_3/target.jpg}
			\imglabel{Bump-3}
		\end{overpic} &
		\includegraphics[width=\resultwidth]{fig7/1_bump_3/good1.jpg} &
		\includegraphics[width=\resultwidth]{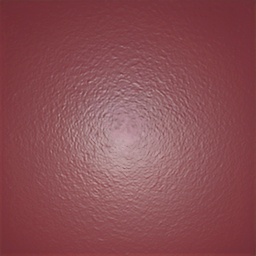} &
		\includegraphics[width=\resultwidth]{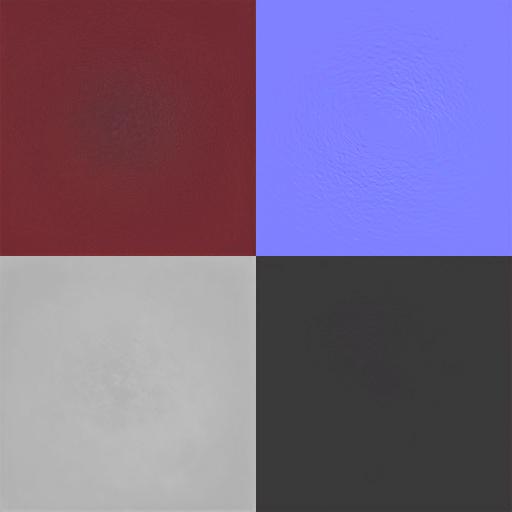} &
		&
		\begin{overpic}[width=\resultwidth]{fig7/1_bump_4/target.jpg}
			\imglabel{Bump-4}
		\end{overpic} &
		\includegraphics[width=\resultwidth]{fig7/1_bump_4/good1.jpg} &
		\includegraphics[width=\resultwidth]{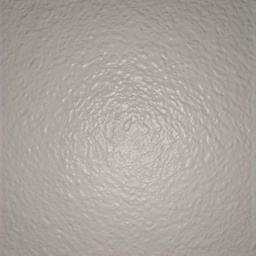} &
		\includegraphics[width=\resultwidth]{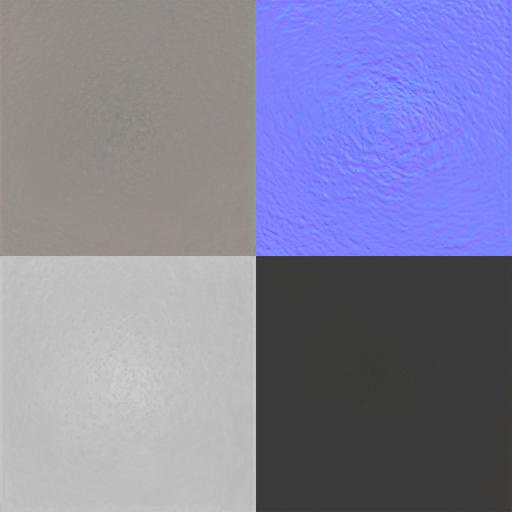}
		\\
		\begin{overpic}[width=\resultwidth]{fig7/2_leather_3/target.jpg}
			\imglabel{Leather-3}
		\end{overpic} &
		\includegraphics[width=\resultwidth]{fig7/2_leather_3/good1.jpg} &
		\includegraphics[width=\resultwidth]{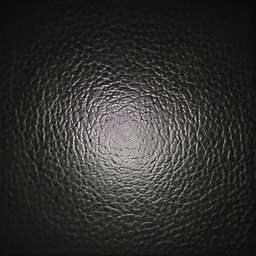} &
		\includegraphics[width=\resultwidth]{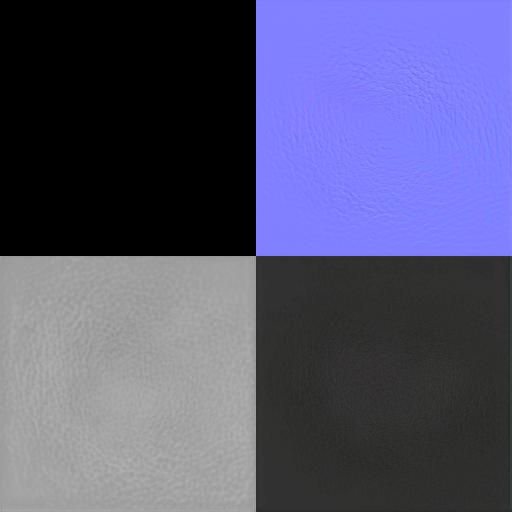} &
		&
		\begin{overpic}[width=\resultwidth]{fig7/2_leather_4/target.jpg}
			\imglabel{Leather-4}
		\end{overpic} &
		\includegraphics[width=\resultwidth]{fig7/2_leather_4/good1.jpg} &
		\includegraphics[width=\resultwidth]{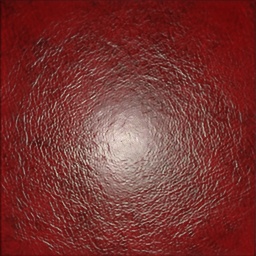} &
		\includegraphics[width=\resultwidth]{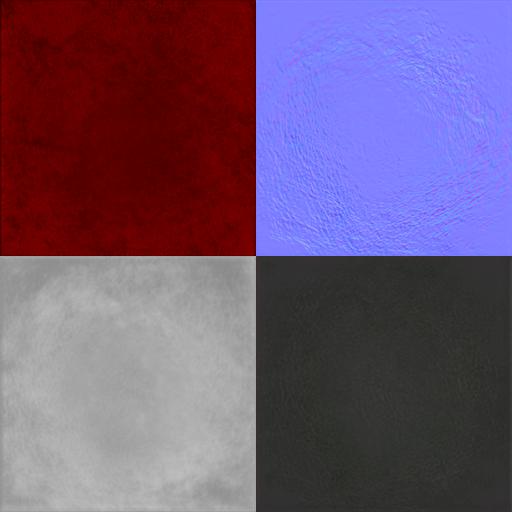}
		\\
		\begin{overpic}[width=\resultwidth]{fig7/2_leather_5/target.jpg}
			\imglabel{Leather-5}
		\end{overpic} &
		\includegraphics[width=\resultwidth]{fig7/2_leather_5/good1.jpg} &
		\includegraphics[width=\resultwidth]{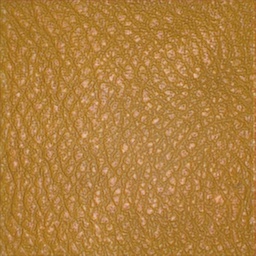} &
		\includegraphics[width=\resultwidth]{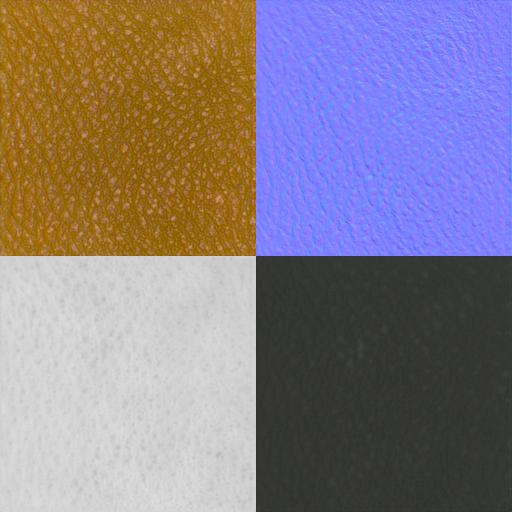} &
		&
		\begin{overpic}[width=\resultwidth]{fig7/2_leather_6/target.jpg}
			\imglabel{Leather-6}
		\end{overpic} &
		\includegraphics[width=\resultwidth]{fig7/2_leather_6/good1.jpg} &
		\includegraphics[width=\resultwidth]{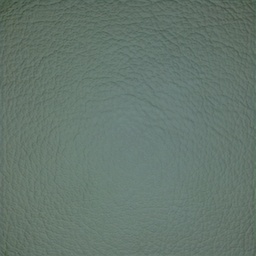} &
		\includegraphics[width=\resultwidth]{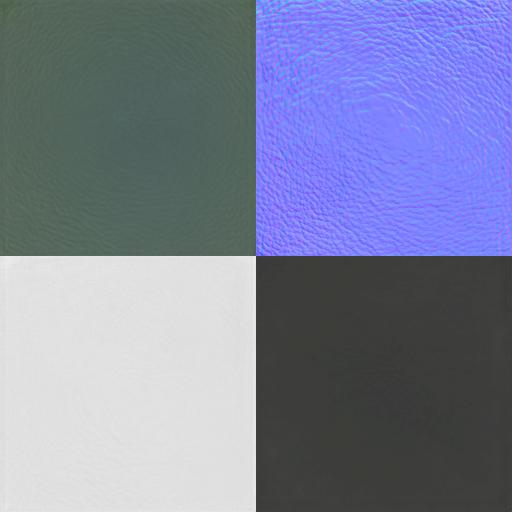}
		\\
		\begin{overpic}[width=\resultwidth]{fig7/3_plaster_3/target.jpg}
			\imglabel{Plaster-3}
		\end{overpic} &
		\includegraphics[width=\resultwidth]{fig7/3_plaster_3/good1.jpg} &
		\includegraphics[width=\resultwidth]{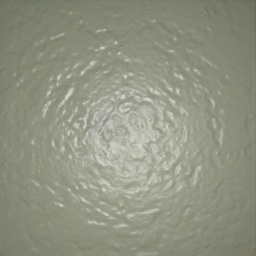} &
		\includegraphics[width=\resultwidth]{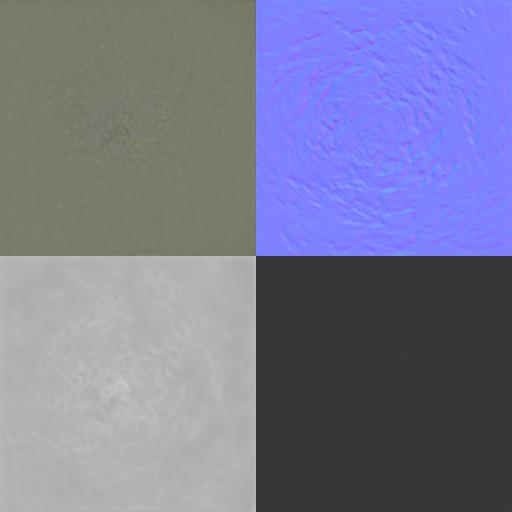} &
		&
		\begin{overpic}[width=\resultwidth]{fig7/3_plaster_4/target.jpg}
			\imglabel{Plaster-4}
		\end{overpic} &
		\includegraphics[width=\resultwidth]{fig7/3_plaster_4/good1.jpg} &
		\includegraphics[width=\resultwidth]{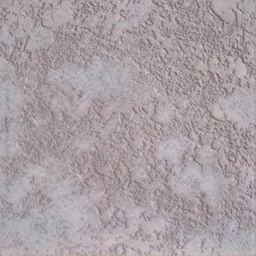} &
		\includegraphics[width=\resultwidth]{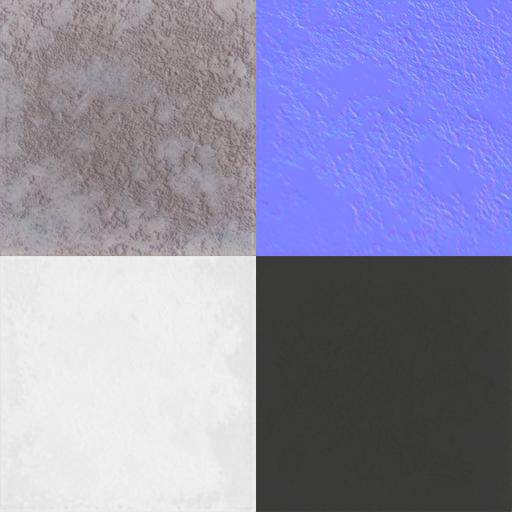}
		\\
		\begin{overpic}[width=\resultwidth]{fig7/4_flake_3/target.jpg}
			\imglabel{Metallicflake-3}
		\end{overpic} &
		\includegraphics[width=\resultwidth]{fig7/4_flake_3/good1.jpg} &
		\includegraphics[width=\resultwidth]{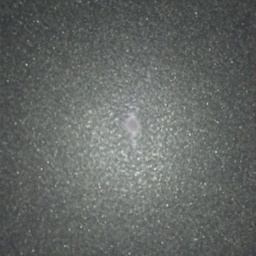} &
		\includegraphics[width=\resultwidth]{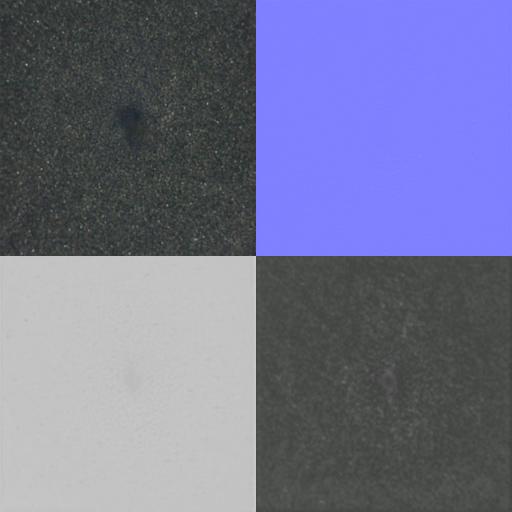} &
		&
		\begin{overpic}[width=\resultwidth]{fig7/4_flake_4/target.jpg}
			\imglabel{Metallicflake-4}
		\end{overpic} &
		\includegraphics[width=\resultwidth]{fig7/4_flake_4/good1.jpg} &
		\includegraphics[width=\resultwidth]{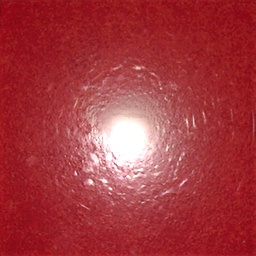} &
		\includegraphics[width=\resultwidth]{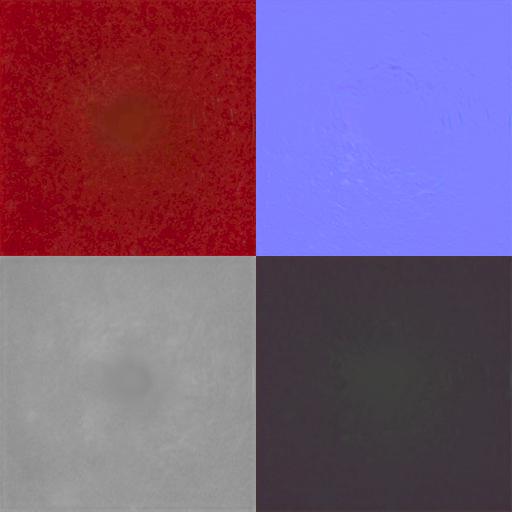}
		\\
		\begin{overpic}[width=\resultwidth]{fig7/5_metal_3/target.jpg}
			\imglabel{Brushmetal-3}
		\end{overpic} &
		\includegraphics[width=\resultwidth]{fig7/5_metal_3/good1.jpg} &
		\includegraphics[width=\resultwidth]{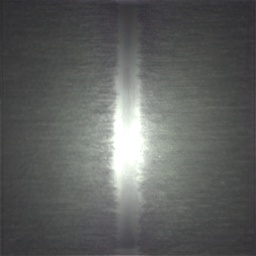} &
		\includegraphics[width=\resultwidth]{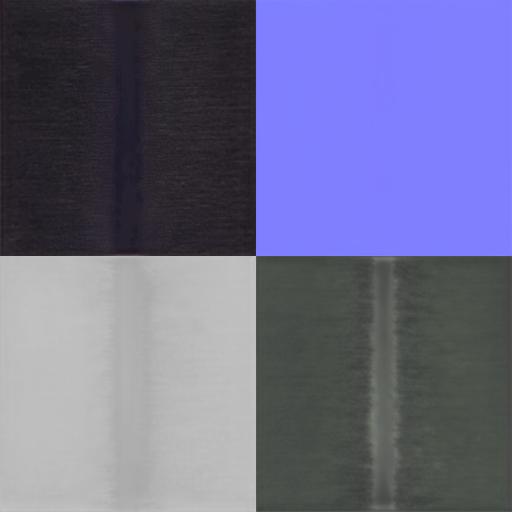} &
		&
		\begin{overpic}[width=\resultwidth]{fig7/6_wood_3/target.jpg}
			\imglabel{Wood-3}
		\end{overpic} &
		\includegraphics[width=\resultwidth]{fig7/6_wood_3/good1.jpg} &
		\includegraphics[width=\resultwidth]{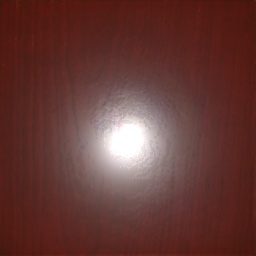} &
		\includegraphics[width=\resultwidth]{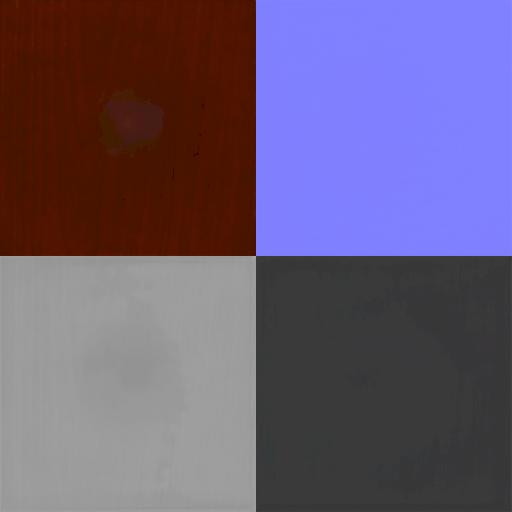}
		\\
		\begin{overpic}[width=\resultwidth]{fig7/6_wood_4/target.jpg}
			\imglabel{Wood-4}
		\end{overpic} &
		\includegraphics[width=\resultwidth]{fig7/6_wood_4/good1.jpg} &
		\includegraphics[width=\resultwidth]{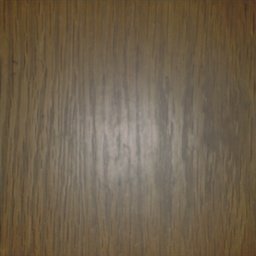} &
		\includegraphics[width=\resultwidth]{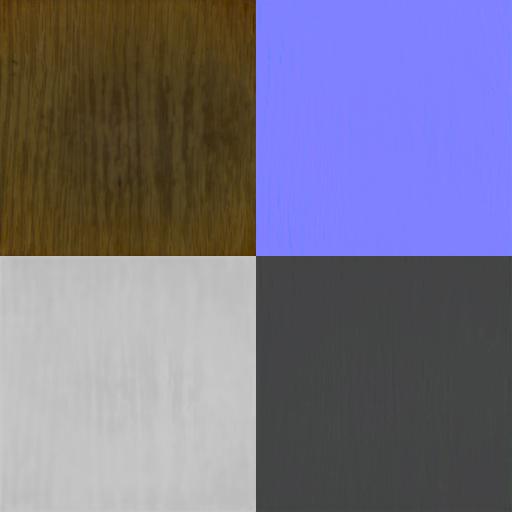} &
		&
		\begin{overpic}[width=\resultwidth]{fig7/6_wood_5/target.jpg}
			\imglabel{Wood-5}
		\end{overpic} &
		\includegraphics[width=\resultwidth]{fig7/6_wood_5/good1.jpg} &
		\includegraphics[width=\resultwidth]{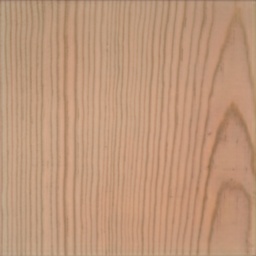} &
		\includegraphics[width=\resultwidth]{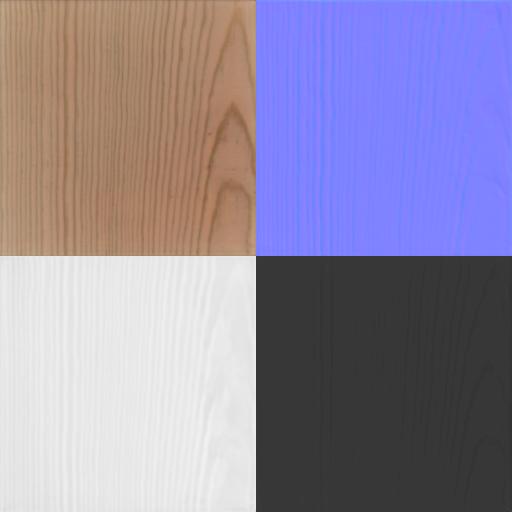}
	\end{tabular}
	\captionsetup{labelfont=bf,textfont=it}
	\caption{\label{fig:Des}
		\textbf{Comparison} to the single input SVBRDF estimation method of Deschaintre et al. \cite{Deschaintre2018}. Due to the nature of the method, their texture patterns are closely aligned with the input image; however, the overall perceptual appearance match is usually worse than our method. In some cases, the method produces specular burn-in, as the strong highlight cannot be fully removed and causes holes in the resulting maps (Plaster-4, Metallicflake-4). Advanced BRDF models like brushed metal and metallic flakes are not explicitly handled by their method and usually fail.
	}
\end{figure*}

\paragraph*{Comparison to neural methods.}
We also compare our method with the forward neural prediction method of Hu et al. \cite{Hu2019}. Their method uses an AlexNet network structure \cite{krizhevsky2012imagenet}, mapping an image of a material sample to the parameters of an appropriate procedural model. We apply their network structure with our BRDFs and lighting conditions, as their original implementation assumes Lambertian materials and outdoor sun/sky lighting. We show the results in Figure \ref{fig:Hu}. In general, we find the method gives moderately accurate results, which moreover tends to become worse for more complex BRDF models and with more parameters. The photo (top) is better matched by our MCMC sampling results (middle) than their prediction (bottom),

To some extent, the method of \cite{Hu2019} is orthogonal to ours, as it can be used as an efficient initialization for our sampling. In Figure \ref{fig:Hu2}, we compare our MCMC sampling results with a random starting point to one using the result of Hu et al. for initialization. This reduces the burn-in period required by the MCMC method.

Finally, we also compare to the single input SVBRDF estimation method of Deschaintre et al. \cite{Deschaintre2018} (See Figure \ref{fig:Des}). This method takes a $256 \times 256$ target image, and produces material maps at the same resolution, pixel-wise aligned to the input. This pixel-wise alignment is not achievable with our method (or any procedural material estimation method). However, the overall perceptual appearance match is usually worse than our method. In some cases, the method produces specular burn-in, as the strong highlight cannot be fully removed and causes holes in the resulting maps (Plaster-4, Metallicflake-4). Advanced BRDF models like brushed metal and metallic flakes are not explicitly handled by their method and usually fail. Finally, their result is fixed at the $256 \times 256$ resolution, and does not support higher resolutions, seamless tiling, nor editability; these benefits come from the use of a procedural model.

Figure~\ref{fig:Comp} shows a quantitative comparison of the Learned Perceptual Image Patch Similarity (LPIPS) metric~\cite{zhang2018unreasonable} between the captured photos and the re-renderings using different methods.

	\section{Conclusion}
\label{sec:conclusion}
%
%
Procedural material models have become increasingly more popular in the industry, thanks to their flexibility, compactness, as well as easy editability.
In this paper, we introduced a new computational framework to solve the inverse problem: the inference of procedural model parameters based on a single input image.

The first major ingredient to our solution is a \emph{Bayesian framework}, precisely defining the posterior distribution of the parameters, combining four components (priors, procedural material model, rendering operator, summary function). The second ingredient is an \emph{Bayesian inference approach} that leverages MCMC sampling to sample posterior distributions of procedural material parameters.  This technique enjoys the generality to handle both continuous and discrete model parameters and provides users additional information beyond single point estimates and allows a cleaner extension to handle discrete parameters.

In the future, we would like to increase the complexity of the models supported even further, to handle materials like woven fabrics, transmissive BTDFs, and more. Finally, extensions to our approach could be used to estimate parameters of procedural models beyond materials, including geometry and lighting, as long as the parameters could be differentiated.

	\bibliographystyle{eg-alpha-doi}
	\bibliography{references}


\end{document}